\documentclass[]{aastex631}

\usepackage{gensymb}
\usepackage{physics}

\def\lsim{\lower.5ex\hbox{$\; \buildrel < \over \sim \;$}}
\def\gsim{\lower.5ex\hbox{$\; \buildrel > \over \sim \;$}}
\def\rg{r_{\rm g}}
\def\nel{n_{{\rm e}^-}}
\def\np{n_{\rm p}}
\def\mel{m_{{\rm e}}}
\def\mp{m_{\rm p}}
\def\sigmat{\sigma_{\rm T}}
\def\msol{\rm{M}_\odot}
\def\mdotsk{\dot{m}_{\rm sk}}
\def\lpsd{l_{\rm{ps}}}
\def\xs{x_{\rm s}}
\def\hs{h_{\rm s}}
\def\xin{x_{\rm in}}
\def\vin{v_{\rm in}}
\def\thetin{\Theta_{\rm in}}
\def\rin{r_{\rm in}}
\def\xo{x_{\rm o}}
\def\mbh{M_{\rm B}}
\def\cot{{\rm{cot}}}
\def\thsk{\theta_{{\rm{sk}}}}
\def\thps{\theta_{{\rm{ps}}}}
\def\cosec{{\rm{cosec}}}
\def\vt{v_{\rm T}}
\def\rsh{r_{\rm sh}}
\def\tg{t_{\rm g}}

\usepackage{comment}
\shorttitle{Colliding shocks in jets}
\shortauthors{Joshi et al.}
\graphicspath{{./}{}}

\begin{document}

\title{Shocks in radiatively driven time dependent, relativistic jets around black holes}

\author[0000-0002-9036-681X]{Raj Kishor Joshi}
\affiliation{Aryabhatta Research Institute of Observational Sciences (ARIES) \\
Manora Peak  \\
Nainital 263001, India  }

\affiliation{Department of Physics, Deen Dayal Upadhyay Gorakhpur University \\
Gorakhpur, 273009, India  \\
}

\author[0000-0002-9851-8064]{Sanjit Debnath}
\affiliation{Aryabhatta Research Institute of Observational Sciences (ARIES) \\
Manora Peak  \\
Nainital 263001, India  }

\author[0000-0002-2133-9324]{Indranil Chattopadhyay}
\affiliation{Aryabhatta Research Institute of Observational Sciences (ARIES) \\
Manora Peak  \\
Nainital 263001, India  }



\begin{abstract}
We study time-dependent relativistic jets under the influence of radiation field of the accretion disk. The accretion disk consists of an inner compact corona and an outer sub-Keplerian disk. The thermodynamics of the fluid is governed by a relativistic equation of state (EoS) for multispecies fluid which enables to study the effect of composition on jet-dynamics. Jets originate from the vicinity of the central black hole where the effect of gravity is significant and traverses large distances where only special relativistic treatment is sufficient. So we have modified the flat metric to include the effect of gravity. In this modified relativistic framework we have developed a new total variation diminishing (TVD) routine along with multispecies EoS for the purpose. We show that the acceleration of jets crucially depends on flow composition. All the results presented are transonic in nature, starting from very low injection velocities, the jets can achieve high Lorentz factors. For sub-Eddington luminosities, lepton dominated jets can be accelerated to Lorentz factors $> 50$. The change in radiation field due to variation in the accretion disk dynamics will be propagated to the jet in a finite amount of time. Hence any change in radiation field due to a change in disk configuration will affect the lower part of the jet before it affects the outer part. 
This can drive shock transition in the jet flow. Depending upon the disk oscillation frequency, amplitude and jet parameters these shocks can collide with each other and may trigger shock cascades.
\end{abstract}


\section{Introduction}
\label{sec:intro}
Astrophysical jets are collimated outflows of plasma associated with various Galatic and extragalactic sources, such as neutron stars, young stellar objects (YSOs), gamma-ray bursts (GRBs), microquasars, and active galactic nuclei (AGNs). These jets span over a wide range of energy and length scales. For example, the jets of YSO travel a distance of several parsecs while AGN jets are observed to remain collimated for hundreds of kiloparsec range. In terms of jet kinetic luminosity, YSO jets have the luminosity of the order $10^{32}\,{\rm erg\,s^{-1}}$ while the GRB jets can reach up to $10^{52}\, {\rm erg\,s^{-1}}$. Since the advent of radio telescopes in the era of modern astronomy, the understanding of these ubiquitous objects has improved significantly, but there is no consensus about the formation, composition, and collimation of these objects \citep{hh06}. Black holes (BH), which sit in the heart of microquasars and AGNs, do not have hard surface and therefore the origin of the jet should be the accreting matter itself. Interestingly, simultaneous radio and X-ray observations show a strong correlation between different spectral states of accretion disks and jets states in microquasars \citep{gfp03,rsfp10, fgr10}, which indicates that the jet originates from the accretion disk. Moreover, the jet is launched from the inner part (less than 100 Schwarzschild radius or $\rg$) of the accretion disk \citep{jbl99, dl12}. While the illusive
corona i.e., the source of hard power law radiation, has been identified as the inner hot part of the accretion disk in many different
models \citep{1976ApJ...204..187S, ct95, 1997MNRAS.288..958G}. Therefore, the compact corona may be responsible for the jet activity. Additionally, the jet which originates from a region very close to the black hole (BH) horizon, will travel through the radiation field of the disk, and should certainly be influenced by
it.       
\\

In recent years, numerical simulations of relativistic jets have played a crucial role in explaining the non-linear dynamics of the system. There are many studies to investigate the propagation of supersonic jets \citep{wamkp14,m19,skr21}. These simulations present a global picture of the jet with a forward bow shock, backflow of the fluid, and multiple shocks within the jet beam. Moreover, the role of magnetic field in the jet formation and collimation has also been studied \citep{fpvj12, kp21,fmgp21}. But the numerical simulations of radiatively driven outflows and jets are very limited, in spite of the fact that the interaction between the radiation field and plasma is not a new subject. The equations of radiation hydrodynamics have been developed by many authors \citep{mm84,k98,p06,r07} and these equations have been used extensively in steady-state investigations of the radiatively driven outflows around compact objects. \cite{i89} used relativistic equation with near disk approximation of radiation field above a Shakura-Sunyaev disk \citep{ss73} to obtain an upper limit of jet terminal speed. \cite{f99} showed that under the influence of radiation arising from disk corona, the jet can achieve significant collimation. So, the radiation not only accelerates the jet, it also helps in collimation. Later, \cite{fth01} showed that electron-proton jets can reach relativistic terminal speeds        
by considering a hybrid disk with inner advection dominated flow (ADAF) \citep{nkh97} and an outer Keplerian disk. The disk model considered by \cite{ct95} with a mixture of matter with sub-Keplerian and Keplerian angular momentum, showed that sub-Keplerian disk can go through a shock transition and create a hot and compact post-shock disk, which can act as an illusive corona.  It may be noted however, inner, puffed hot disk acting like a corona has been suggested by other authors too \citep{1976ApJ...204..187S,1997MNRAS.288..958G}. Numerical simulations of advective disks show that the extra thermal gradient term in hot post-shock region automatically generates the bipolar outflows \citep{dcnm14,lcksr16} which can be accelerated by the radiation field of the disk.     
In the relativistic radiation hydrodynamic regime, investigations by \cite{c05} and \cite{vc17,vc18,vc19}  
clearly highlight that the radiation field of the accretion disk plays a significant role in the acceleration of the jets. In addition to these steady-state investigations there are limited numerical simulations in the non-relativistic regime \citep{cc02a,csnr12,rvc21,jcl22}. To study the effect of radiation on winds there are some simulations of line driven winds \citep{psk2000,ybl18}, but the line driven force is only effective when the wind is cold, so the line driven force is not applicable in hot plasma which is being considered in this paper.\\

In this paper, we have made an attempt to understand the dynamics of the jet in the time-dependent radiation field of the accretion disk. The steady-state investigations by  \cite{vc17,vc18,vc19} and \cite{ftrt85}
have shown that the radiation field can produce internal shocks in jets but how do these shocks evolve under a time-dependent radiation field is not known yet. So, the central theme of this paper is to investigate the formation and evolution of these shocks in relativistic jets due to its interaction with the radiation field of the accretion disk. We study jets which are launched with subsonic speeds close to the central BH. Since the jets are launched close to the BH so one has to consider the effect of gravity at the jet base.
However, due to the huge length scales jets traverse, in most parts
gravity is negligible and special relativity suffices. Previous studies like \cite{ftrt85,vkmc15} used the Newtonian or pseudo-Newtonian potential as an additive  source term for gravity in the special relativistic Euler equation but this destroys the conservative nature of equations of motion which is essential
in the Eulerian upwind numerical schemes we are using. Hence, we include the gravity by modifying the special relativistic metric itself and thereby retaining the conservative form of the equations of motion. The presence of gravity makes the jet transonic, while we avoid general relativistic details. The temperature of the jet varies by a large value so we use a relativistic equation of state (EoS) in our analysis given by \cite{cr09}. This EoS (abbreviated as CR) handles the variation of the adiabatic index and also allows us to study the effect of plasma composition on the jet solution. 
\\

The internal shocks in the jet are identified as bright knots in the jets. These shocks are the sites for particle acceleration in the jets, resulting in non-thermal emissions. The jet models with shock are very popular and have been used to explain the outbursts in microquasars \citep{kss2000}, AGNs \citep{wamkp14}, as well as GRB jets \citep{rm94}. Shock very close to the jet base around a stellar mass BH was considered to explain the high-energy power law emission \citep{lrwc11}.
Numerical simulations of jets show that the shocks form in the jet beam whenever a fast moving supersonic plasma catches up with a slower moving flow ahead. 
It is well known that the radiation field accelerates by pushing the jet material. However, due to relativistic effects, radiation would also decelerate the jet close to the disk via a process called radiation drag \citep{i89,f99,cc02,cdc04,c05}. Therefore if the disk radiation is time-dependent, can it accelerate the jet at lower altitudes to trigger shocks by crashing the lower altitude faster blobs onto the slower portion of the jet above it? As these jet shocks are time-dependent, how does the shock strength evolve in time? If the disk is oscillating, can the resulting time-dependent radiation field produce
multiple shocks?
Further, would these shocks collide with each other and generate shock cascades?
How would the jet solutions depend on the baryon fraction of the jet?
In this paper, we address these questions.

In section \ref{sec:equations} we present the assumptions and governing equations used in this paper. In section \ref{sec:method} we outline the methodology to obtain the solutions. Section \ref{sec:results} is devoted to explain and discuss the key findings of this paper. In the end, in section \ref{sec:concl} we conclude our work. 

\section{Assumptions and governing equations} \label{sec:equations}
We study the jets driven by the radiation field of an accretion disk around a non-rotating black hole. 
We do not include the accretion disk directly in our study, the disk plays a supportive role by supplying the radiation field. All the disk parameters used in our study are required only for the calculation of the  radiative moments. We have not explored the direct connection between disk and jet generation, instead, we assume the jet injected with a finite subsonic initial velocity and some initial temperature.
In this study we restrict our calculations for a relativistic non-rotating ($u_\phi=0$), axis-symmetric ($\partial/\partial\phi=0$), on-axis ($\theta=0$) jet. We assume a conical geometry of the jet with a narrow opening angle. The energy-momentum tensor for jet material ($T^{\alpha\beta}_M$) and radiation field ($T^{\alpha\beta}_R$) are given as
 
\begin{equation}
T^{\alpha\beta}_M= \rho h u^\alpha u^\beta +p g^{\alpha\beta};\,\,\, T^{\alpha\beta}_R= \int I_{\nu}l^{\alpha}l^{\beta} d\nu d\Omega
\label{eq:em_tensor}
\end{equation}      

where $u^{\alpha}$ represent the components of four velocity, $l^{\alpha}$s are the direction cosines, $I_\nu$ is the specific intensity of the radiation field. $d\Omega$ is the solid angle subtended by the field point on the jet axis to the source point on the accretion disk. $\rho$ is the rest mass density of fluid and $h$ is the specific enthalpy of the fluid which is related to the internal energy density $e$ and pressure $p$ as 

\begin{equation}
h=\frac{(e+p)}{\rho}
\label{eq:enthalpy}
\end{equation}     

The space-time metric is given by 

\begin{equation}
ds^2=-g_{tt}dt^2+dr^2+r^2d\theta^2+r^2sin^2\theta d\phi^2
\label{eq:metric}
\end{equation}
Where $g_{tt}$ is given as;

\begin{equation}
g_{tt}=\left(1+\Phi\right); \,\,\, \Phi=-\frac{1}{2(r-1)}
\label{eq:gtt}
\end{equation}

We have used the geometric unit system where $2G=\mbh=c=1$, $\mbh$ is the mass of the BH, $G$ is the gravitational constant, and $c$ is the speed of light. The information of gravity is supplied through $g_{tt}$, somewhat in the spirit of weak-field approximation, although the $\Phi$ is given by the Paczy\'nsky-Wiita potential \citep{pw80}, instead of the Newtonian one. 
This way, we retain the conservative nature of equations of motion.
In this unit system, the Schwarzschild radius ($\rg=2G \mbh/c^2$) is the unit of length and $\tg=2G\mbh/c^3$ is the unit of time. 

The equations of motion for relativistic radiation hydrodynamics have been derived before \citep{p06, r07}, here we present only a brief description to obtain the equations of motion. The continuity equation which represents the conservation of mass flux is given as
$(\rho u^\alpha)_{;\alpha}=0$.
The equations of motion are essentially the conservation of energy momentum tensor or, $(T^{\alpha\beta}_M)_{;\beta}=-(T^{\alpha\beta}_R)_{;\beta}=G^\alpha$.

For a conical jet along the axis of symmetry implies that $u^r$ is the only significant component of the four-velocity and conservations of mass flux and energy-momentum tensor can be written as a set of three conservation laws:

\begin{equation}
\frac{\partial D}{\partial t}+\frac{1}{r^2}\frac{\partial}{\partial r}\left(r^2\sqrt{g_{tt}}Dv\right)=0
\label{eq:cont_cons}
\end{equation}

\begin{equation}
\frac{\partial M}{\partial t}+\frac{1}{r^2}\frac{\partial}{\partial r}\left(r^2\sqrt{g_{tt}}(Mv+p)\right)+E\frac{\partial \sqrt{g_{tt}}}{\partial r}-\frac{2\sqrt{g_{tt}}p}{r}=G^r
\label{eq:mom_cons}
\end{equation}

\begin{equation}
\frac{\partial E}{\partial t}+\frac{1}{r^2}\frac{\partial}{\partial r}\left(r^2\sqrt{g_{tt}}(E+p)v\right)+(E+p)v\frac{\partial \sqrt{g_{tt}}}{\partial r}=G^t
\label{eq:eng_cons}
\end{equation}

Here $v$ is the three velocity of the jet. The four velocity ($u^r$) can be written in terms of $v$ as
$u^r=\gamma v$, where $\gamma$ is the Lorentz factor given as $\gamma^2=-u_tu^t=1/(1-v^2)$. The conserved quantities $D,\,M,\,{\rm and}\,E$ are the mass density, momentum density, and total energy density in the laboratory frame given as \citep[also see][]{rcc06,skrhc21}

\begin{eqnarray}
\label{eq:conservq1}
D=\gamma \rho \\
\label{eq:conservq2}
M= \gamma^2 \rho h v \\
E= \gamma^2 \rho h -p 
\label{eq:conservq}
\end{eqnarray}   

$G^r$ and $G^t$ are the components of the radiation force, which under the present set of assumptions are explicitly given as
\citep{mm84,k98,p06}

\begin{equation}
G^t=\frac{\gamma}{\sqrt{g_{tt}}}vR^r,~\&~ G^r=\gamma^3 R^r
\label{eq:gtfinal}
\end{equation} 
where
\begin{equation}
R^r=\rho_e\frac{\sigmat}{\mel c}\left[(1+v^2)F_{\rm rd}-v(E_{\rm rd}+P_{\rm rd})\right]=\rho_e\left[(1+v^2)\mathcal{F}-v(\mathcal{E}+\mathcal{P})\right]
\label{eq:radterm}
\end{equation}

Here $\rho_e$ is the total lepton density and $\mathcal{F}={\sigmat F_{\rm rd}}/({\mel c}),\,\mathcal{E}={\sigmat E_{\rm rd}}/({\mel c}),\,\mathcal{P}={\sigmat P_{\rm rd}}/({\mel c})$, also, $E_{\rm rd},~F_{\rm rd},~\&~P_{\rm rd}$ are the zeroth moment or energy density, first moment or the flux, and second moment or the pressure of the radiation field, respectively. \\

To solve the equations of motion (\ref{eq:cont_cons}-\ref{eq:eng_cons}) we need an additional closure relation which relates the thermodynamic variables $h, p$, and $\rho$. This closure relation is known as the equation of state (EoS). In this study, we consider an EoS for relativistic multispecies fluid proposed by \cite{cr09} (abbreviated as CR EoS). CR EoS is a very close fit to the exact one given by \cite{c39}. The CR has already been used for different astrophysical problems \citep{camrs14,sc19,scl20,jcry21,jcl22}. 

The EoS is given as 
\begin{equation}
e=\rho f
\label{eq:eos}
\end{equation}

where,
\begin{equation}
f=1+(2-\xi)\Theta\left[\frac{9\Theta+6/\tau}{6\Theta+8/\tau}\right]+\xi\Theta\left[\frac{9\Theta+6/\eta\tau}{6\Theta+8/\eta\tau}\right]
\label{eq:eos2}
\end{equation}
In equations (\ref{eq:eos}, \ref{eq:eos2}) $\rho$ is the mass density of fluid given as $\rho=\Sigma_i n_im_i=\nel \mel (2-\xi+\xi/\eta)$, where $\xi=\np/\nel$, $\eta=\mel/\mp$ and
$\nel$, $\np$, $\mel$ and $\mp$ are the electron number
density, the proton number density, the electron rest mass, and proton rest mass. $\Theta=p/\rho$ is a measure of temperature and $\tau=2-\xi+\xi/\eta$. The expression for sound speed is given as

\begin{equation}
a^2=\frac{1}{h}\frac{\partial p}{\partial \rho}=-\frac{\rho}{Nh}\frac{\partial h}{\partial \rho}=\frac{\Gamma\Theta}{h}
\label{eq:soundsp}
\end{equation}
Here, $N$ is the polytropic index given as

\begin{eqnarray}
N =  \rho \frac{\partial h}{\partial p}-1=\frac{\partial f}{\partial \Theta}
 =6\left[(2-\xi)\frac{9\Theta^2+24\Theta/\tau+8/\tau^2}{(6\Theta+8/\tau)^2}\right]+6\xi \left[\frac{9\Theta^2+24\Theta/(\eta \tau)+8/(\eta \tau)^2}{\{6\Theta+8/(\tau \eta)\}^2}\right]
\label{eq:poly}
\end{eqnarray}

The polytropic index is a function of temperature. Hence, we do not need to supply it as a free parameter. From equation (\ref{eq:poly}) one can see that for higher values of temperature $\Theta>>1$, $N \rightarrow 3$, and $N \rightarrow 3/2$ for $\Theta<<1$. The adiabatic index is related with the polytropic index as
\begin{equation}
\Gamma=1+\frac{1}{N}
\label{eq:adindx}
\end{equation}  

\subsection{Steady state equations of motion}

In the steady-state, we impose ${\partial }/{\partial t}\equiv 0$ on equations of motion (\ref{eq:cont_cons}-\ref{eq:eng_cons}).
Integrating equation (\ref{eq:cont_cons}) then leads to the mass-outflow rate.

\begin{equation}
\dot{M}=\rho\gamma v\mathcal{A}
\label{eq:mdot}
\end{equation}
where $\mathcal{A}(\propto r^2)$ is the cross-section of the jet.
The energy balance equation or 1$^{\rm st}$ law of thermodynamics
is given by $u_\mu(T^{\mu \nu}_M)_{;\nu}=-u_\mu(T^{\mu \nu}_R)_{;\nu}$ which gives the temperature gradient 
\begin{equation}
\frac{d \Theta}{d r}=-\frac{\Theta}{N}\left[\frac{2}{r}+\frac{\gamma^2}{v}\frac{dv}{dr}\right]
\label{eq:dthdr}
\end{equation} 
Integrating the above and combining with equation \ref{eq:mdot} gives the adiabatic relation
\citep[also, see][]{kscc13,jcl22}
\begin{equation}
\dot{\mathcal{M}}=\gamma v r^2 \Theta^{3/2}(3\Theta+4/\tau)^{k_1}(3\Theta+4/\eta\tau)^{k_2}{\rm exp}(k_3)
\label{eq:scriptm}
\end{equation}

where
\begin{equation}
k_1=\frac{3}{4(2-\xi)},\, k_2=\frac{3\xi}{4},\, k3=-\frac{3}{\tau}\left[\frac{2-\xi}{3\Theta+4/\tau}+\frac{\xi}{\eta(3\Theta+4/\eta\tau)}\right]    
\end{equation} 
Since interaction of radiation with the jet in the Thompson scattering regime is isentropic, so the entropy-outflow rate is a constant along a streamline in the steady-state, but discontinuously jumps up in case there is a shock transition in the jet. The relativistic Euler equation [$(\delta^r_\mu+u^ru_\mu)(T^{\mu \nu}_M+T^{\mu \nu}_R)_\nu=0$] can be simplified to the following form,
\begin{equation}
\gamma^2\frac{dv}{dr}=\frac{{2a^2}/{r}-\frac{1}{2(1+\Phi)}\frac{d\Phi}{dr}+{\gamma R^r}/{\rho h}}{v-\frac{a^2}{v}}
\label{eq:std_eul}
\end{equation}
where, the last term in the numerator $\frac{\gamma R^r}{\rho h}$ is the radiation momentum deposition term, which can be simplified to the form
\begin{equation}
\frac{\gamma R^r}{\rho h}=\frac{\gamma(2-\xi)}{\tau h}\left[(1+v^2)\mathcal{F}-v(\mathcal{E}+\mathcal{P})\right]
\label{eq:rad_mom_deps}
\end{equation}
We can integrate equations (\ref{eq:std_eul}) and (\ref{eq:dthdr}) to obtain another constant of motion for steady flow known as the generalized relativistic Bernoulli parameter (${\cal B}_{\rm e}$).

\begin{equation}
{\cal B}_{\rm e}=-hu_t{\rm exp}(-X); \,\,\, X=\int \frac{\gamma R^r}{\rho h} dr
\label{eq:genbp}
\end{equation}

\begin{figure}
\gridline{\fig{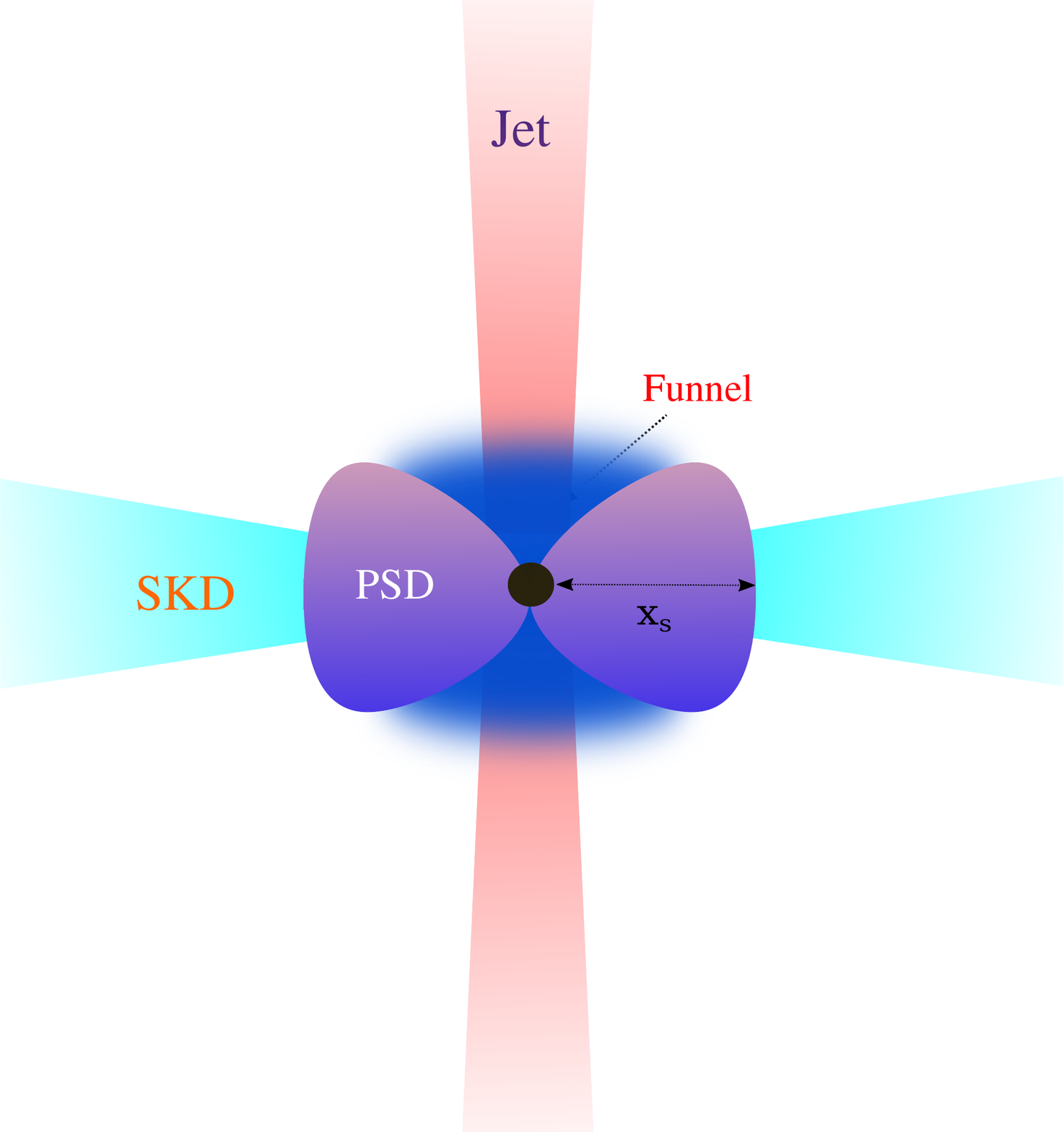}{0.4\textwidth}{(a)}
          \fig{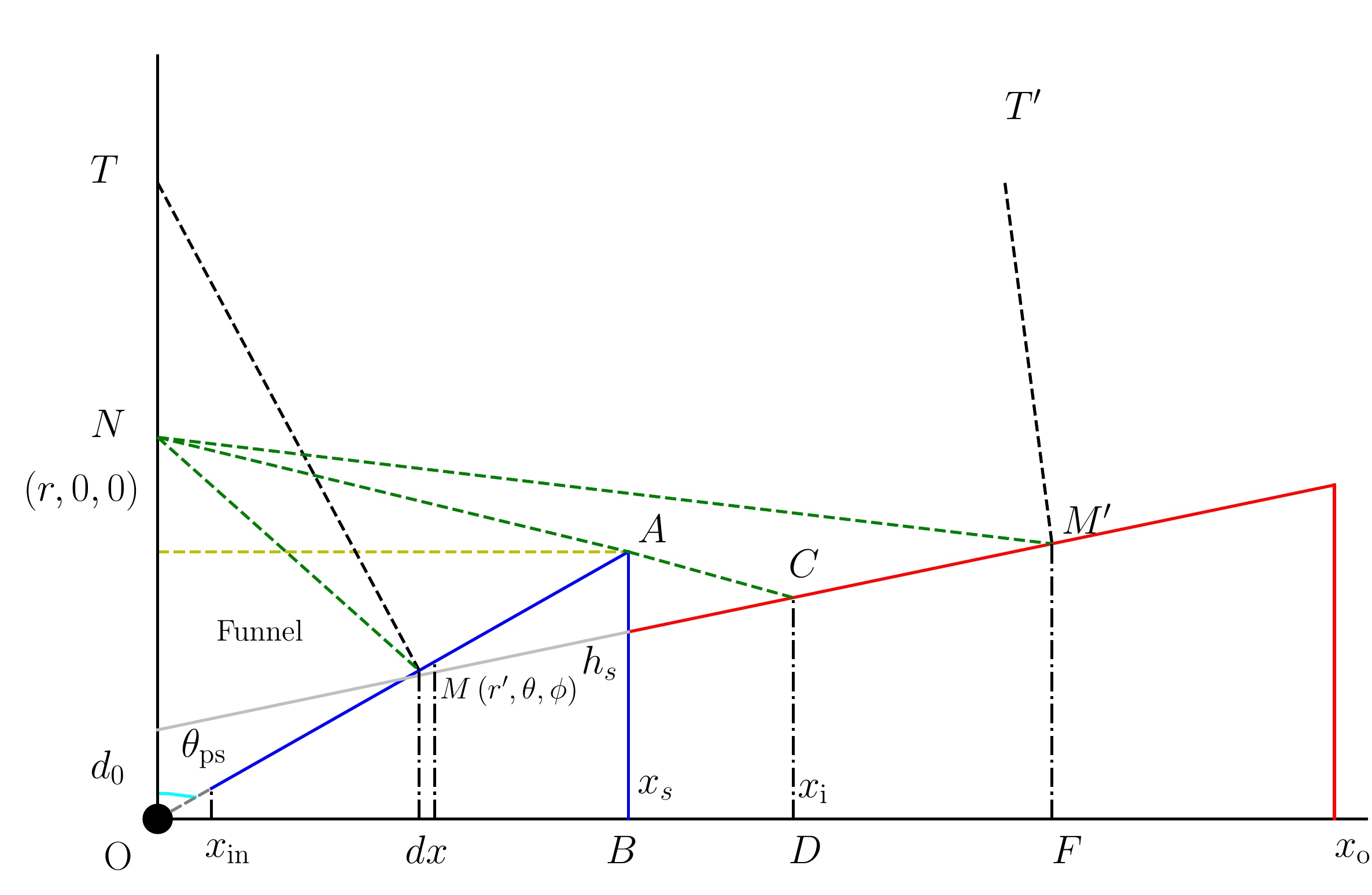}{0.6\textwidth}{(b)}}
\caption{(a) Schematic representation of the disk-jet system. The shock location ($\xs$), and various components of disk-jet system are shown. The blue region marks the funnel above the PSD. (b)- Cartoon diagram to show the cross sectional view of the disk. The blue portion is inner compact corona. SKD is shown with solid red lines. N is the field point on jet axis where radiative moments are calculated. M and ${\rm M^\prime}$ are the source point on the corona and SKD.}
\label{fig:cartoon}
\end{figure}

We consider an advective type accretion disk \citep{f87,c89,lrc11,2017MNRAS.469.4221K} for this study. Figure \ref{fig:cartoon}(a) shows a typical cartoon diagram for the disk and jet system. As mentioned earlier, the disk plays only a supportive role in our calculations by supplying the radiation field. Depending upon the accretion rate advective disks have moderate to low radiative efficiency \citep{scl20}. The disk has two components, an inner puffed up post shock disk called PSD and a sub-Keplerian disk (SKD). The sub-Keplerain part of the disk supplies the soft photons and the geometrically thick PSD also known as corona supplies the hard photons. This type of disk structure can mimic hard to hard-intermediate spectral states in microquasars. The PSD terminates at the horizon but we take the inner edge of the corona at $1.5\,\rg$ as the disk is expected to emit negligible radiation from a region within it. Numerical simulations of advective accretion flows suggest that the SKD is flatter than the PSD \citep{lcksr16} and the ratio of height to the horizontal extent of corona can vary from 1.5 to 10, so we take a semi-vertical angle $\theta_{\rm sk}=85^{\rm o}$ for SKD and intercept of SKD on jet axis is taken to be $d_0=0.4h_{\rm s}$, where $\hs$ is the height of corona taken as $\hs=2.5\xs$. The outer edge of SKD is taken at $\xo=3500$. The details of obtaining the radiative moments are presented in \cite{cc00,c05,vkmc15}, here we present only a brief description. \\

In Fig.\ref{fig:cartoon}(b) we have shown a cross sectional view of disk and jet sytem. $M$ and $M^\prime$ are the source points on PSD and SKD, respectively. $TM$ and $T^\prime M^\prime$ are the local normals on disk surface. We calculate the radiative moments on the field point $N$. The inner edge of SKD is at $\xs$ but due to the shadow effect \citep{c05}, the observer at $N$ sees the SKD starting from $x_{\rm i}$ given as

\begin{equation}
x_i (r)=\frac{r-d_0}{(r-\hs)/\xs+{\rm cot}\theta_{\rm sk}}
\label{eq:skd_in_edge}
\end{equation}

We need the intensities of various disk elements to obtain the radiative moments of respective disk components. In SKD, we assume the synchrotron emission \citep{st83} to be the dominant emission mechanism.     

\begin{equation}
I_{\rm sk}=\left[\frac{16}{3}\frac{e^2}{c}\left(\frac{eB_{\rm sk}}{\mel c}\right)^2\Theta_{\rm sk}^2n_{\rm sk}\right]\frac{\left(d_0\,{\rm sin\theta_{sk}}+x\,{\rm cos\theta_{sk}}\right)}{3}\,{\rm erg\, cm^{-2} s^{-1}}    
\end{equation}

where $B_{\rm sk},\,\Theta_{\rm sk},\,n_{\rm sk},\,x$ represent magnetic field, local dimensionless temperature, electron number density, and horizontal distance from central object respectively. We assume a stochastic magnetic field in SKD with a constant magnetic to gas pressure ratio $p_{\rm mag}=\beta{p_{\rm gas}}$.
The general definitions of radiative moments are 
\begin{equation}
 E_{\rm rd}=\frac{1}{c}\int I d\Omega,~~ F^i_{
 \rm rd}=\int I l^i d\Omega, \mbox{ and } P^{ij}_{\rm rd}=\frac{1}{c}\int I l^i l^j d\Omega.
 \label{eq:moments}
\end{equation}
and at each field point (e. g. N) these quantities have to be computed from both the sources SKD and PSD. All the components of the radiation moments should ideally be numerically computed in equations \ref{eq:num_mom}. The details of computing these numerically are given in \citet{vkmc15} and also in Appendix \ref{sec:app1} of this paper. 

\begin{figure}[!h]
 \centerline {\includegraphics[width=14cm,height=14cm, keepaspectratio]{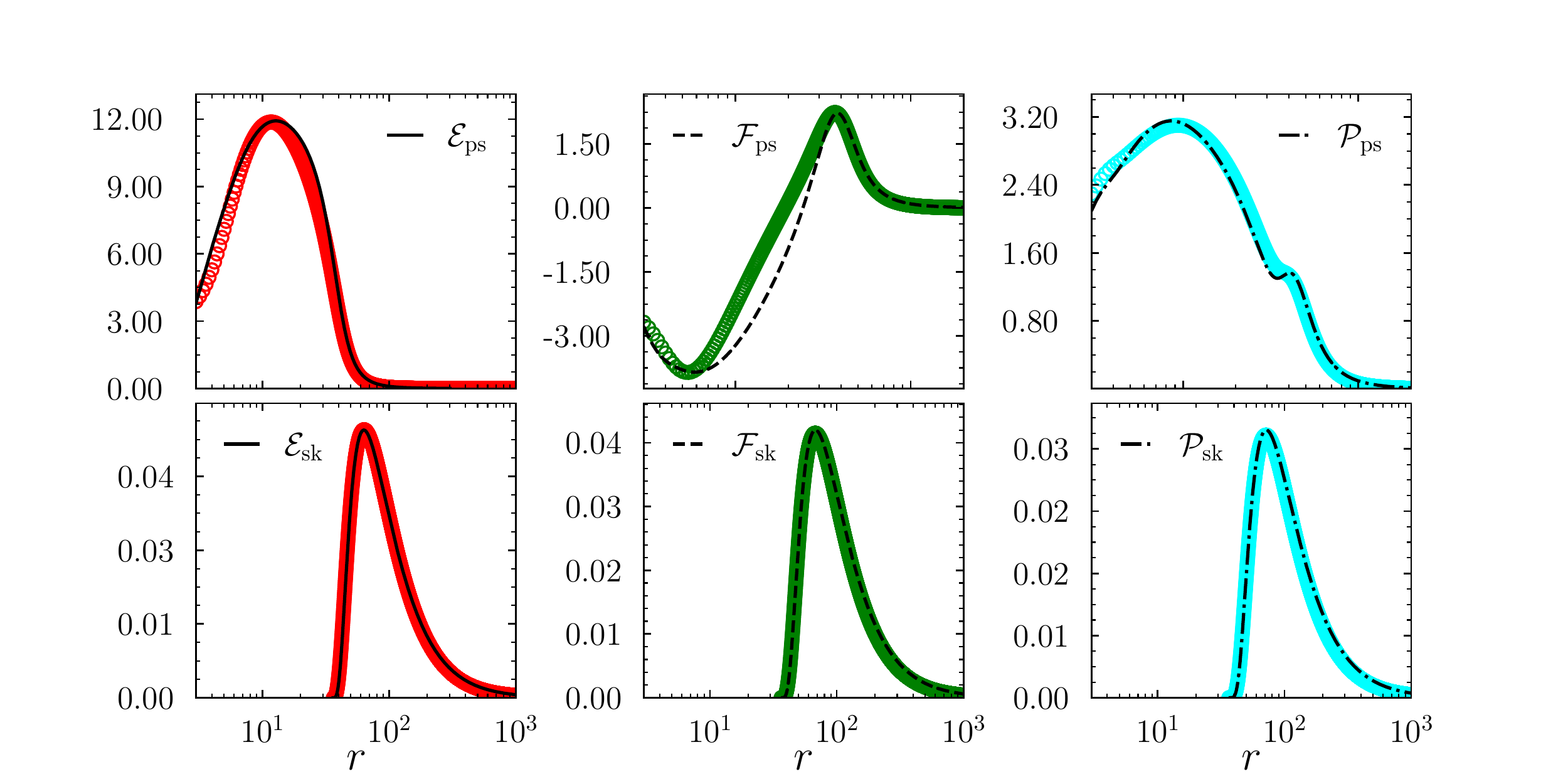}} 
  \caption{The radiative moments from PSD and SKD (excluding the constant factors). Lines (solid, dashed, dashed-dotted) represent the correct numerically integrated moments and open circles (red, green and cyan) represent
  moments obtained by fitting functions.}
  \label{fig:Rad_moms}
\end{figure}

However, numerically computing the radiative moments in the simulation code at every cell and at every time step would slow down the code significantly. So we fitted the numerically computed radiative moments (equations \ref{eq:skd_ez}-\ref{eq:psd_pz}) with approximate analytical functions (equations \ref{eq:eng_alb}-\ref{eq:pres_alb} for SKD and \ref{eq:eng_alb_ps}-\ref{eq:pres_alb_ps} for PSD in Appendix \ref{sec:app2}) which depend on disk parameters like $\xs, ~\hs ~\& ~ \xin$. The sub-Keplerain accretion rate controls the position of $\xs$ (equation \ref{eq:xs_mdot}) and luminosities (equation \ref{eq:skd_lum_func}-\ref{eq:psd_lum_func}) of the disk components. The comparison between the moments obtained from numerical integration (solid, dashed, dashed-dotted lines) and the analytical fitted functions (red, green and cyan open circles) for $\xs=15$ is shown in Fig.(\ref{fig:Rad_moms}). The fitting is good, and the advantage is, any time variation of $\mdotsk$ will produce a time dependent radiation field. But the moments
are given as analytical functions and need not be numerically computed at every cell and at every time step. It may be noted that,
${\cal E},~{\cal F}~\mbox{and}~{\cal P}$
obtained from equation \ref{eq:eng_alb}--\ref{eq:pres_alb}, and \ref{eq:eng_alb_ps}--\ref{eq:pres_alb_ps} for a given $\mdotsk$ are used in equation \ref{eq:radterm}. 

\section{Method of obtaining solutions}
\label{sec:method}

\subsection{Steady state solutions}
The steady-state solutions are obtained by simultaneously integrating equations (\ref{eq:std_eul}) and (\ref{eq:dthdr}). The jet at its base is hot and slow and therefore subsonic. The jet is powered by the thermal gradient term and the radiative term against gravity and becomes supersonic at some value of $r$. In other words, jets are transonic in nature and at some point $r_c$ called sonic point, the flow velocity $v_c$ crosses the local sound speed $a_c$. At $r_c$, $dv/dr\rightarrow0/0$ gives the sonic point conditions

\begin{equation}
v_c=a_c
\label{eq:spcond1}
\end{equation}  

\begin{equation}
\frac{2a_c^2}{r_c}-\left(\frac{1}{2(1+\Phi)}\frac{d\Phi}{dr}\right)_c+\left(\frac{\gamma R^r}{\rho h}\right)_c=0
\label{eq:spcond2}
\end{equation}

From equation (\ref{eq:spcond2}) we obtain the temperature ($\Theta_c$) at the sonic point and $dv/dr$ is obtained by using ${\rm L'H\hat{o}pital's}$ rule. We integrate the equations (\ref{eq:std_eul}) and (\ref{eq:dthdr}) inwards and outwards starting from sonic points to obtain the solutions. The steady-state solutions provides the injection values for the simulation code.

\subsection{Time dependent solutions and numerical setup}
We perform the time-dependent calculation using a relativistic total variation diminishing (TVD) routine.
TVD scheme introduced by \cite{h83} is a second-order, Eulerian, finite difference scheme, which was initially proposed to solve the set of hyperbolic, non-relativistic hydrodynamic equations. However, many relativistic numerical codes have been built based on the same philosophy \citep[see][for details]{mm03, mm15}. A relativistic TVD code to incorporate CR EoS has been built before \citep{rcc06, crj13}. The detailed description of the code is presented in \citep{rcc06}, here we present only brief details. \\

The set of equations (\ref{eq:cont_cons})-(\ref{eq:eng_cons}) can be written in the form

\begin{equation}
\frac{\partial \vb*{q} }{\partial t}+\frac{1}{r^2}\frac{\partial (r^2 \vb*{F})}{\partial r}=\vb*{S}
\label{eq:rhd_conservative}
\end{equation}
with the state and flux vectors given as

\begin{equation}
\vb*{q}=\begin{pmatrix} D \\ M \\ E	 \end{pmatrix},\,\,\,\,\, \vb*{F}=\begin{pmatrix} \sqrt{g_{tt}}D v \\ \sqrt{g_{tt}} (Mv+p) \\ \sqrt{g_{tt}}(E+p)v \end{pmatrix}
\label{eq:state_vect}
\end{equation}

and the source term $\vb*{S}$ as

\begin{equation}
\vb*{S}=\begin{pmatrix} 0 \\ G^r-E\frac{\partial \sqrt{g_{tt}}}{\partial r}+\frac{2\sqrt{g_{tt}}p}{r} \\ G^t-(E+p)v\frac{\partial \sqrt{g_{tt}}}{\partial r}	 \end{pmatrix}
\label{eq:source}
\end{equation}     

The set of equations \ref{eq:rhd_conservative} is solved in two steps. In the first step, the state vectors $\vb*{q}$ at the cell center $i$ at a given time step $n$ are updated using the modified fluxes $\bar{f}_{i+1/2}$ calculated at cell boundaries.

\begin{equation}
\vb*{L}\vb*{q}^{n}=\vb*{q}^{n-1}-\frac{\Delta t^n}{\Delta x}\left(\bar{f}_{i+1/2}-\bar{f}_{i-1/2}\right)
\end{equation}
where,
\begin{equation}
 \bar{f}_{i+1/2} = \frac{1}{2}\left[\vb*{F}^n_i 
+\vb*{F}^n_{i+1}\right]-\frac{\Delta x}{2\Delta t^n}
\sum_{k=1}^3\beta_{k,i+1/2}\vb*{R}_{k,i+1/2}^n,
\end{equation}
where,
\begin{equation}
 \beta_{k,i+1/2} = Q_k\left(\frac{\Delta t^n}{\Delta x}\lambda_{k,i+1/2}^n
+\gamma_{k,i+1/2}\right)\alpha_{k,i+1/2}-\left(g_{k,i}+g_{k,i+1}\right).
\end{equation}
Here, 
$$
\alpha_{k,i+1/2} = \vb*{L}_{k,i+1/2}^n\cdot\left(\vb*{q}_{i+1}^n
-\vb*{q}_i^n\right)
$$

Explicit forms of eigenvalues $\lambda_{k,i}$, eigenvectors $\vb*{L}~\&~\vb*{R}$, flux limiters $g_{k,i}$, $\gamma_{k,i+1/2}$, $Q_k$ are given in \cite{rcc06}. In the second step, the radiation and curvature terms are added as source term      
in the updated values of state vectors. The next step consists of recovering the updated primitive variables ($\rho, v, p$) from the updated conserved variables $D,~M,~E$ which is done by inverting the equations \ref{eq:conservq1}-\ref{eq:conservq}. These equations along with  EoS can be expressed as a polynomial of $\gamma$, given as 

\begin{equation}
\sqrt{\gamma^2-1}D+g_1\gamma\left[\frac{(2-\eta\tau)}{2(1-\eta)}\left(\frac{9\gamma g_1+6D\sqrt{\gamma^2-1}/\tau}{3\gamma g_1+4D\sqrt{\gamma^2-1}/\tau}\right)+\eta\frac{(\tau-2)}{2(1-\eta)}\left(\frac{9\eta\gamma g_1+6D\sqrt{\gamma^2-1}/\tau}{3\eta\gamma g_1+4D\sqrt{\gamma^2-1}/\tau}\right)+1\right]-M=0
\label{eq:lofac_polyn}
\end{equation}
where $g_1=\gamma M-E\sqrt{\gamma^2-1}$. We solve equation \ref{eq:lofac_polyn} using Newton-Raphson method to obtain $\gamma$, and once $\gamma$ is known the primitive variables can be obtained as 
\begin{equation}
v=\frac{\sqrt{\gamma^2-1}}{\gamma}
\label{eq:vcal}
\end{equation}

\begin{equation}
\rho=\frac{D}{\gamma}
\label{eq:rhocal}
\end{equation}

Since $v$, $\gamma$, $E$ and $M$ are known, then using equations \ref{eq:conservq2} and \ref{eq:conservq}, the pressure $p$ is obtained as the root of a cubic equation 

\begin{eqnarray}
a_1p^3+a_2p^2+a_3p+a_4=0, \\ \mbox{where, }
a_1=27\eta, \,\, a_2=9\left[\frac{4}{\tau}(1+\eta)\rho-\eta(E-Mv)\right] \nonumber \\ 
a_3=\frac{12\rho}{\tau}\left[\frac{2\rho}{\tau}+(1+\eta)\rho-\eta(E-Mv)\right], \,\,\, 
a_4=\frac{16\rho^2}{\tau^2}[\rho-E+Mv] \nonumber
\label{eq:pres_recov}
\end{eqnarray}

We employ a continuous inflow boundary condition at the injection cell and outflow boundary conditions at the outer boundary using ghost cells. In this time-dependent study, we want first to obtain the steady state solutions as is predicted by theoretical analytical investigations. And then we want to study the influence of accretion disk oscillation on the jet dynamics, while keeping the injection parameters same as the steady state values. 

\subsubsection{Time dependent disk as observed by an inertial observer on the axis}
\label{sec:disc_inertial}
\begin{figure}[!h]
 \centerline {\includegraphics[width=15cm,height=6cm]{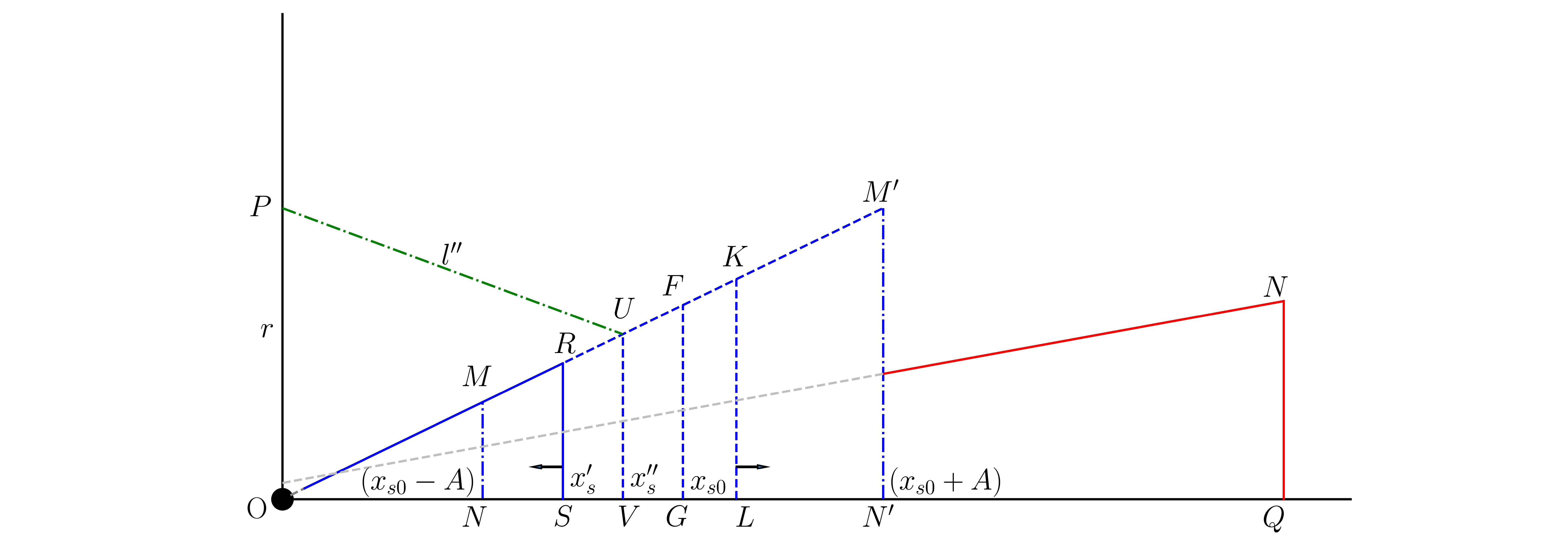}} 
  \caption{Schematic representation of time-dependent disk configuration as seen by an observer P on jet axis. $x_{s0}$ is the mean position of shock and $\xs^\prime$ is the actual position of shock at time t.}
  \label{fig:retdpos}
\end{figure}

In this paper, the time dependence of the jet is not imposed through
the injection parameters, but by making the radiation field from the accretion disk time-dependent. In microquasars, quasi-periodic oscillations (QPO) are observed in hard X-rays associated with the oscillation of the inner part of the accretion disk \citep{ndmc12}. If there is a fluctuation of the accretion rate $\mdotsk$ at the outer edge of the disk, the fluctuation would travel inwards. The time variability of the accretion rate $\mdotsk$
is given in equation \ref{eq:accretvar}, which induces a sinusoidal
variation in the size of PSD or $\xs$, is of the form,
\begin{equation}
\xs^\prime=x_{\rm s0}+A\,{\rm sin} (2\pi f_{\rm s} t)
\label{eq:shockt}
\end{equation} 
where the oscillation is about a mean $x_{s0}$ (distance OG fo Fig. \ref{fig:retdpos}) with an amplitude of $A$ (NG=GN$^\prime$), as shown in Fig.(\ref{fig:retdpos}) and frequency $f_{\rm s}$. The information about the disk-fluctuation propagates at the speed of light to the observer P, i. .e, it takes a finite amount of time to reach the jet axis. Hence the different points on the jet axis see a different disk configuration \citep[see][]{jcl22}.
In other words, the radiation emitted from $\xs^{\prime\prime}$ at $t$ reaches $P$ after a time $t+\delta t$ ($\delta t=l^{\prime\prime}/c$, see Fig. \ref{fig:retdpos}), and in the same time-interval the PSD boundary will move from $\xs^{\prime\prime}$ to $\xs^\prime$. Therefore,

\begin{equation}
\delta t=l^{\prime\prime}/c=\sqrt{{\xs^{\prime\prime}}^2+(r-\hs^{\prime\prime})^2}  \mbox{;  in units of } c=1
\label{eq:delta_t}
\end{equation}

The position $\xs$ is updated after each time interval $dt$ which is determined by the TVD code.
So we can write
\begin{equation}
\xs^{\prime \prime}-\xs^{\prime}=s dt  
\label{eq:pos_diff}
\end{equation}
Where $s=0.5(v_{{\rm s},\,t}+v_{{\rm s},\,t+dt})$ is the average velocity in the time interval between $t$ to $t+dt$. The instantaneous velocities can be obtained from equation \ref{eq:shockt},

\begin{equation}
 v_{\rm s}=A2\pi f_{\rm s} {\rm cos}(2\pi f_s t)
 \label{eq:instvelsok}
\end{equation}

From equations (\ref{eq:delta_t}) and (\ref{eq:pos_diff})
we can write
\begin{equation}
(\xs^{\prime\prime}-\xs^{\prime})^2=s^2\left[{\xs^{\prime\prime}}^2+(r-h_s^{\prime\prime})^2\right]    
\label{eq:ret_pos}
\end{equation}

We solve equation (\ref{eq:ret_pos}) at each $r$ and at every time step to obtain the $\xs$ as seen by the observer there, and the corresponding radiative moments are calculated using the analytical estimates of the radiative moments (equations \ref{eq:eng_alb}-\ref{eq:pres_alb_ps}). 

\section{Results}
\label{sec:results}
\subsection{Code verification} 
\begin{figure}
\centerline {\includegraphics[width=12cm,height=12cm, keepaspectratio]{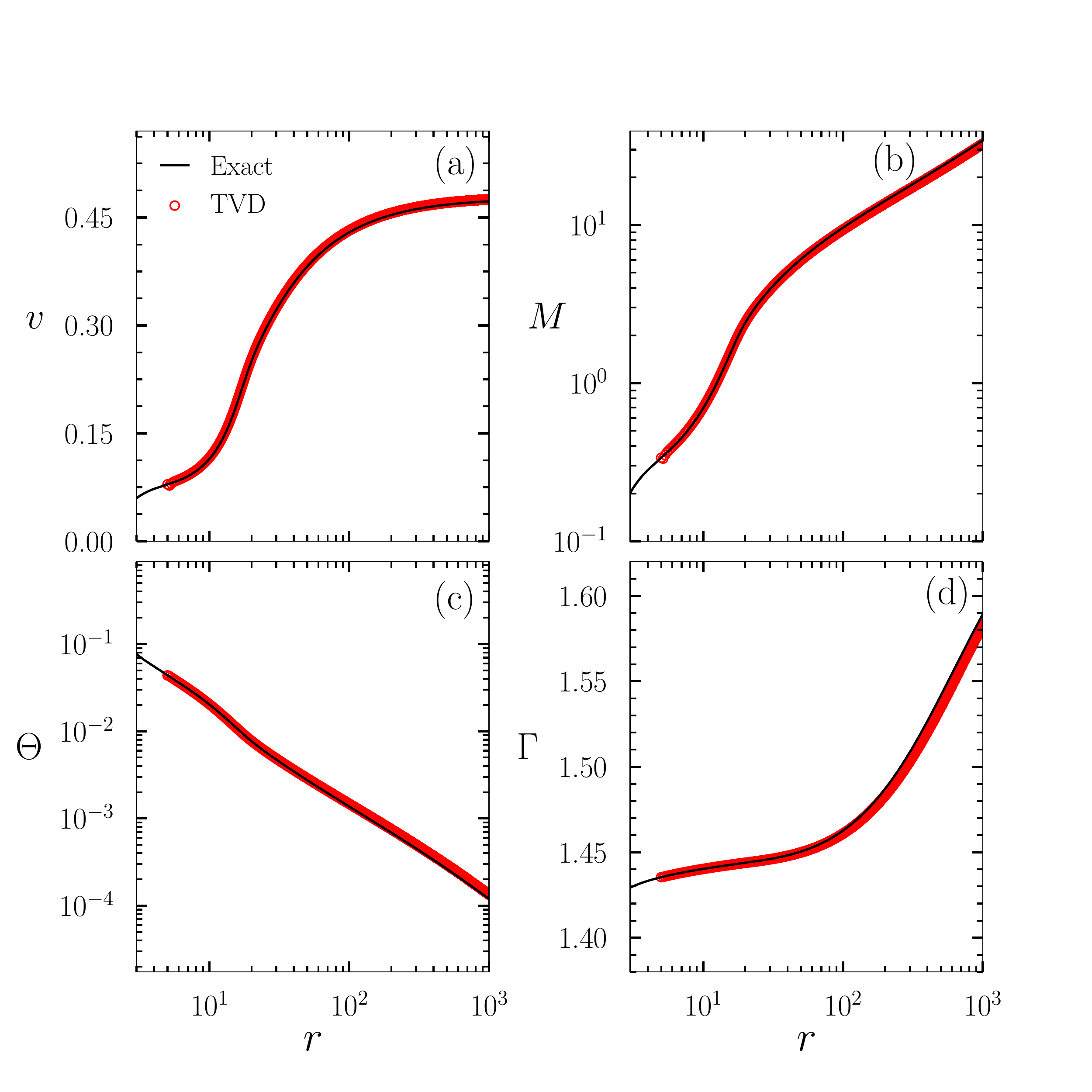}} 
\caption{Comparison between the steady state solutions obtained from TVD code (red open circles) and analytical method (solid black lines), for $v_{\rm in}=0.065,\,\,\Theta_{\rm in}=0.037,\,\, r_{\rm in}=5.0$. The jet is electron-proton or $\xi=1.0$ fluid.}
\label{fig:steady_comp}
\end{figure}

The relativistic Riemann problem is a great tool to test the time-dependent hydrodynamic code, and the special relativistic TVD code with CR EoS was indeed tested
against the analytical relativistic Riemann problem \citep[see][]{jcry21}.
To check the correctness and accuracy of the code in the relativistic radiation hydrodynamic regime in presence of gravity, presently we compare the numerical result with the steady-state semi-analytical radiatively driven relativistic jet solution. In Fig. \ref{fig:steady_comp} we have shown the solutions obtained from the TVD code marked by open red circles and analytical solutions shown by solid black lines. The evolution of various variables like flow velocity $v$, Mach number $M~(\equiv v/a)$, $\Theta$ which is the measure of temperature, and adiabatic index $\Gamma$ along the jet is shown in panels (a)-(d), respectively. The SKD accretion rate $\mdotsk$ is taken to be 10, which produces a PSD with $\xs=8.212$ and disk luminosity is $l=1.0$. The SKD accretion rate $\mdotsk$ and disk luminosity $l$ are in the units of the Eddington accretion rate ($\dot{M}_{\rm Edd}=1.44\times10^{17}M_{\rm{B}}/\msol \,{\rm g/s}$) and the Eddington luminosity ($\dot{L}_{\rm Edd}=1.26\times10^{38}M_{\rm{B}}/\msol \, {\rm erg /s}$) respectively. The injection parameters with which the jet is launched are chosen from the analytical solution and are given by $v_{\rm in}=0.065,\,\,\Theta_{\rm in}=0.037,\,\, r_{\rm in}=5.0$. The sonic point of the steady-state solution corresponding to these parameters is at $r_c=14.0$. We divide the computational domain of total length $1000 r_g$ into uniform 6000 cells. It may be noted that the simulation of outflows requires higher resolution compared to accretion disk
simulations, because of the presence of sharp gradients at outflow injection boundaries. The comparison shows that the code regenerates the solution very accurately. The adiabatic index of the jet at the base is moderately relativistic, but becomes non-relativistic at large $r$, which suggests that the use of CR EoS to describe the thermodynamics of the jet is appropriate. 

\subsection{Effect of disk luminosity on jet solution}

\begin{figure}
\centerline {\includegraphics[width=13cm,height=13cm, keepaspectratio]{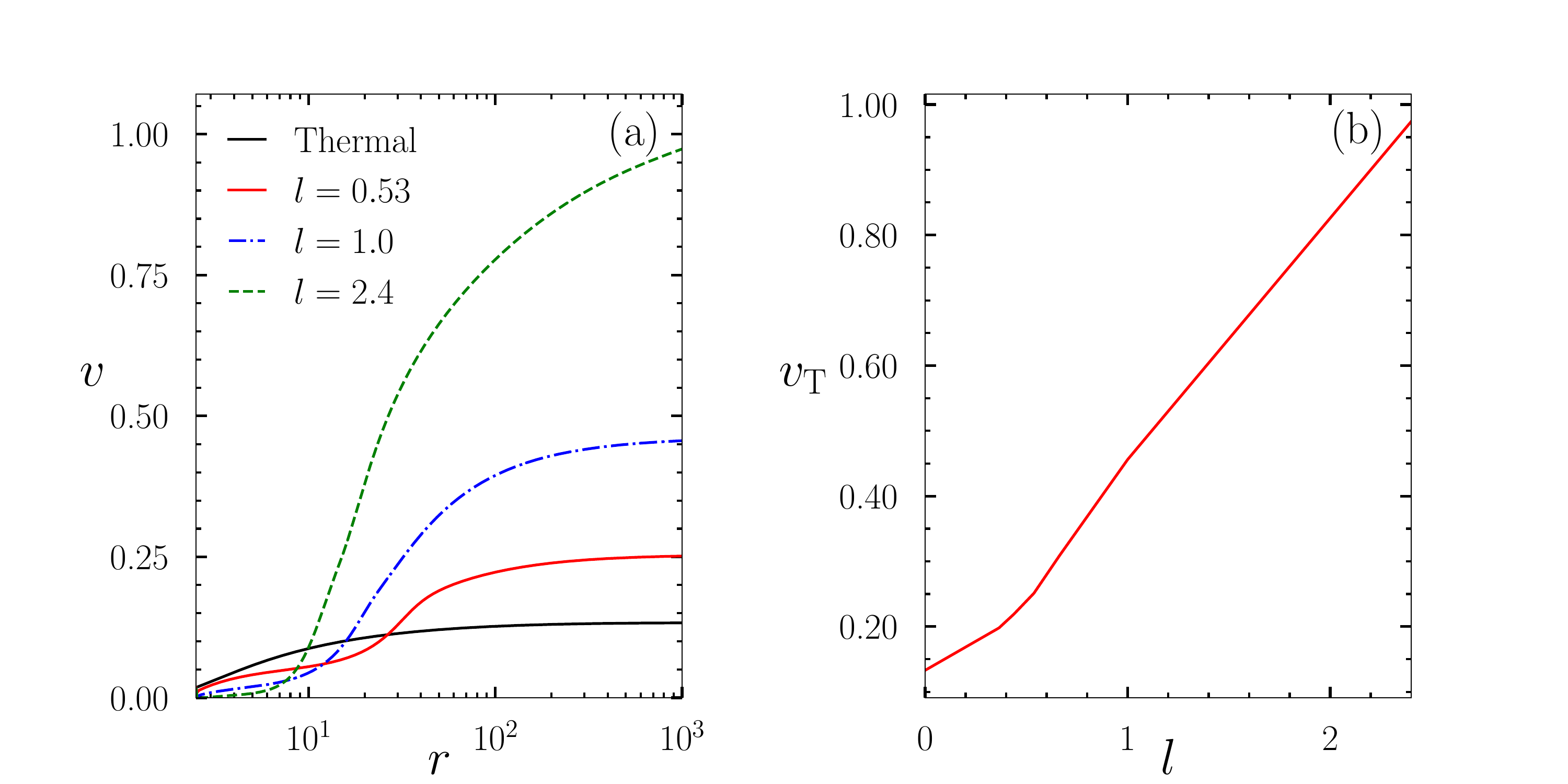}} 
\caption{The effect of radiation on the jet solutions. Velocity profiles for various solutions with different disk luminosities are plotted in panel(a) and panel (b) shows the terminal velocities.}
\label{fig:lum_comp}
\end{figure}

In Fig. \ref{fig:lum_comp}a, we compared solutions of electron-proton ($\xi=1$) jets corresponding to different disk luminosities, all starting with the same injection parameters   
$\vin=0.018,\,\thetin=0.063,\,\rin=2.5$. These injection parameters are taken from a thermally driven steady-state solution characterized by the sonic point location $r_c=15.0$.
Each curve represents thermally driven jet (solid, black curve), jets driven by disk luminosities $l=0.53$ (solid, red), $l=1.0$ (dash-dot, blue) and $l=2.4$ (dashed, green), all in units of Eddington luminosity. The terminal speed of the thermally driven jet is $v_{\rm T}=0.13c$. It is expected that higher disk luminosity would
accelerate the jets to higher $v_{\rm T}$. And indeed Fig. \ref{fig:lum_comp}b shows, $v_{\rm T}$
increases with $l$, to the extent that an electron-proton jet for sub-Eddington disk luminosity ($l<1$) may reach
mildly relativistic speeds ($v_{\rm T}\sim 0.4-0.5c$),
while for the super-Eddington luminosities the jet can achieve relativistic terminal velocities. Interestingly, higher $l$
suppresses the jet velocity $v$ in the funnel like region above
the PSD.
From Fig. \ref{fig:Rad_moms}, it is clear that the radiative
flux ${\cal F}_{\rm ps}<0$ in the funnel like region between the
inner walls of the PSD. So with higher $l$, the ${\cal F}_{\rm ps}$ will actually suppress the jet speed within the funnel region, although at higher altitude would push the jet. Therefore $\vt$ increases with $l$.

\subsection{Effect of jet composition}
\label{sec:comp_effect}
\begin{figure}[!h]
\centerline {\includegraphics[width=13cm,height=13cm, keepaspectratio]{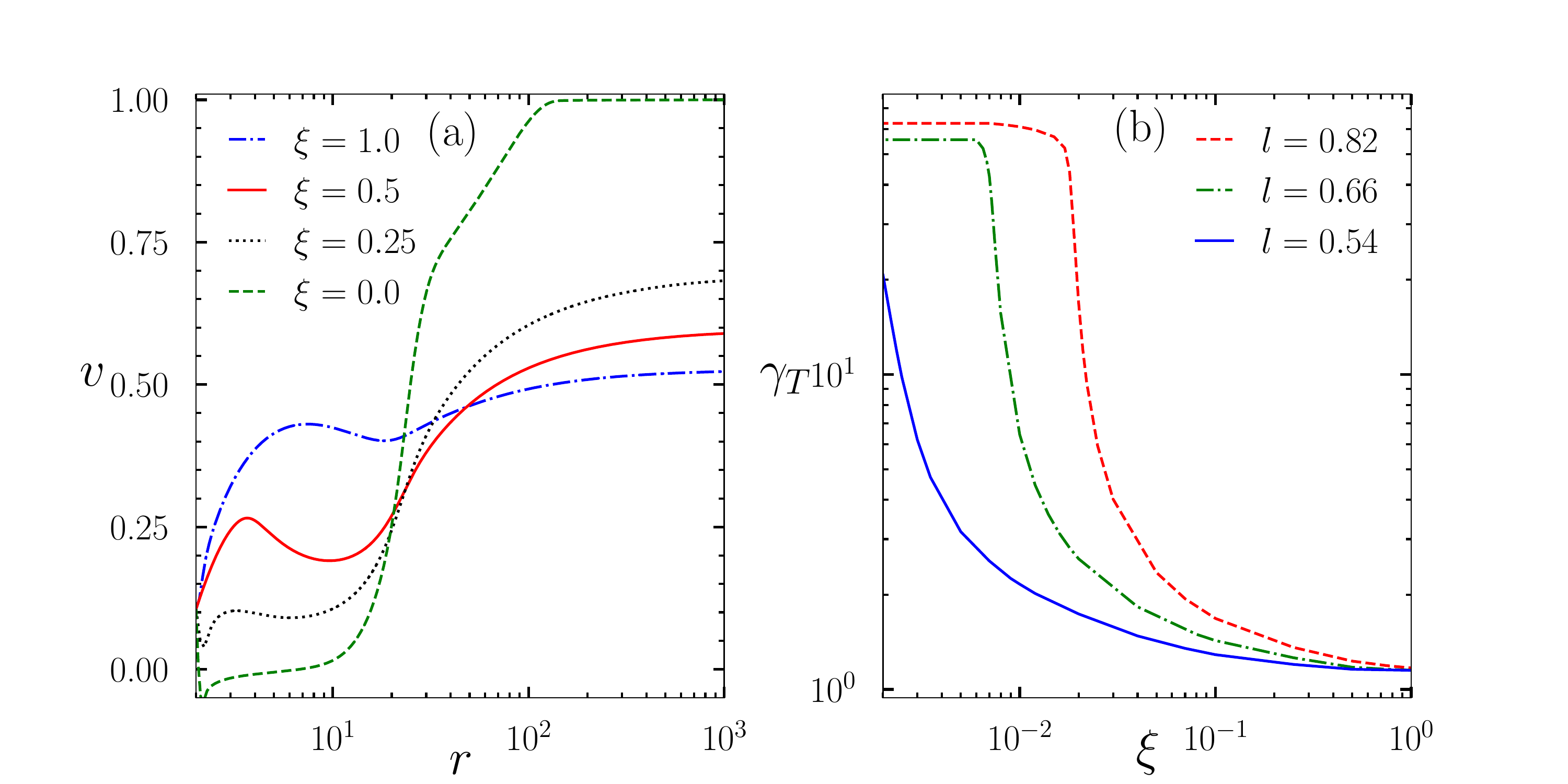}} 
\caption{(a)-The velocity profiles for different plasma compositions of the jet. Panel (b) shows the variation of terminal Lorentz factor with respect to composition parameter for various disk luminosities.}
\label{fig:comp_effect}
\end{figure}

The interaction between the jet plasma and radiation is dominated by Thompson scattering, so the amount of momentum transferred to the jet will be higher for the lepton dominated jet which can be clearly seen from the equation (\ref{eq:radterm}). To study the effect of composition we fix injection and disk parameters and the solutions are obtained for various values of $\xi$. In Fig. (\ref{fig:comp_effect} a) we have plotted the velocity profiles of various jet solutions, with the injection parameters
$\vin=0.11,\, \thetin=0.22, \, \rin=2.0$. The disk luminosity is $l=0.82$. The radiation term $R^r$ for solutions with higher fraction of protons (higher $\xi$) will be weaker resulting in lower terminal speeds. But the radiation term inside the funnel is negative so the lower magnitude of $R^r$ means the low value of deceleration, so the proton dominated flows turn out to be faster at lower $r$. So the electron-proton jet ($\xi=1.0$), plotted with dash-dotted blue line is fastest near the jet base. The composition parameter significantly affects the terminal velocities which can be clearly seen from Fig.(\ref{fig:comp_effect})(a). The terminal velocity for electron-positron flow ($\xi=0$), plotted with dashed-green line is $v_{\rm T}=0.999876$ and the corresponding terminal Lorentz factor is $\gamma_{\rm T}=63.51$. In the right panel we have plotted the variation of terminal Lorentz factor $\gamma_T$ with respect to the jet composition for luminosities $l=0.82$ (dashed-red), $l=0.66$ (dash-dotted blue), and $l=0.54$ (solid-blue).   
These results clearly suggest that the interaction of disk radiation with the jet plasma can play a crucial role in the acceleration of the jet. In contrast to the electron-proton jets, the lepton dominated jets can reach ultra-relativistic speeds even for the sub-Eddington luminosities.     

\subsection{Jet solutions under the time dependent radiation field}
\begin{figure}[!h]
\centerline {\includegraphics[width=24cm, height=11cm]{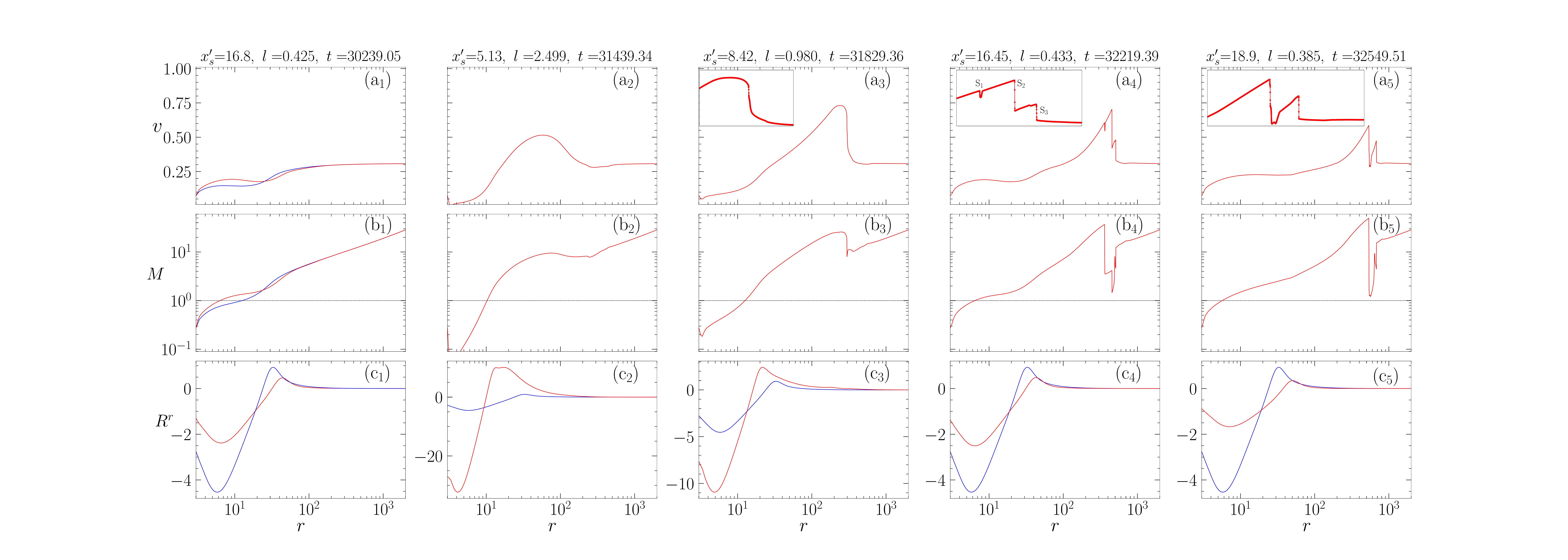}}
\caption{The evolution of jet solution under the influence of disk oscillations. The variation of velocity (${\rm a_1}-{\rm a_5}$), Mach number (${\rm b_1}-{\rm b_5}$), radiative term (${\rm c_1}-{\rm c_5}$) is plotted at different epochs. The steady state solution (blue, solid) is obtained for steady luminosity $l=0.597$ and steady PSD configurtation at $x_{{\rm s}0}=12$. The PSD oscillation is for amplitude $A=7$, time period of $T=2000\tg$ about an equilibrium value $x_{{\rm s}0}$ of the size of PSD.}
\label{fig:Shockcolsn}
\end{figure}

\begin{figure}[!h]
\centerline {\includegraphics[width=20cm, height=10cm]{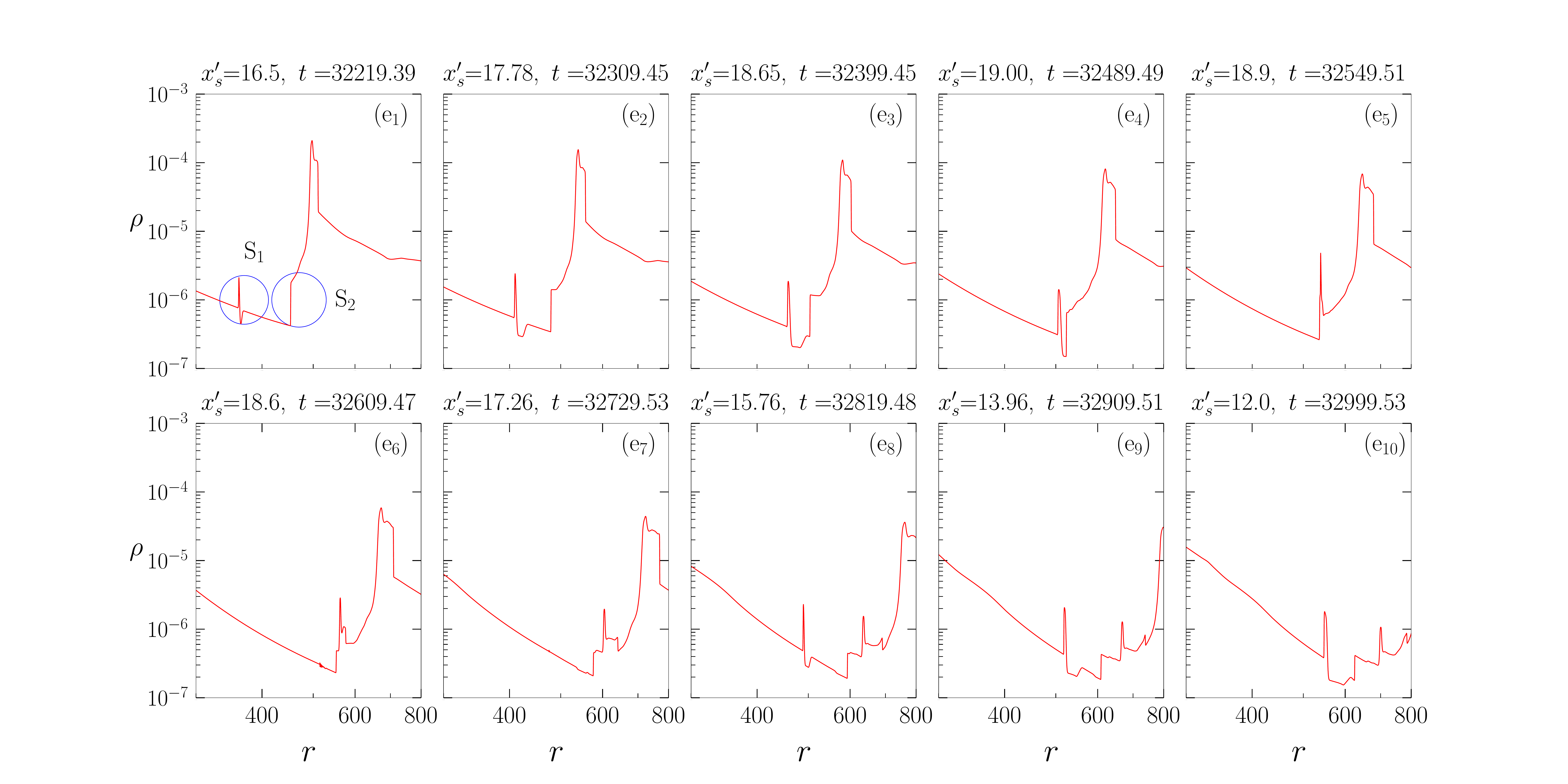}} 
\caption{Density distribution $\rho$ of the jet at vaious time steps are plotted to the collision between shock $\rm S_1$ and $ \rm S_2$ in panels $\rm (e_1-e_5)$. Panels $\rm (e_6-e_{10})$ show the $\rho$ after the collision. The PSD oscillation is same as in Fig. \ref{fig:Shockcolsn}.}
\label{fig:collision}
\end{figure}

To study the effect of disk oscillation on jet dynamics, we first obtain a steady-state solution corresponding to injection parameters of steady-state analytical solution. The steady-state time is taken to be $t=30000\,t_g$. In Figs. \ref{fig:Shockcolsn}
\& \ref{fig:collision} the injection parameters are
$\vin=0.076,\,\,\,\thetin=0.064,\,\,\,\rin=3.0$.
The disk luminosity for the steady-state configuration is $l=0.597$ and the outer edge of the PSD is at $\xs=12.0$ (obtained from equation \ref{eq:xs_mdot}). The velocity ($v$), Mach number ($M$) and net radiation force term ($R^r$) for the steady-state solutions are plotted (blue, solid curve) in Figs. \ref{fig:Shockcolsn} ($\rm a_1$), ($\rm b_1$), ($\rm c_1$), respectively. The sonic point for steady-state solution is at $r_c=13.0$. Once the steady-state solution is obtained we induce a variability in
$\mdotsk$ (equation \ref{eq:accretvar}) such that it imposes an oscillation on $\xs$ with the amplitude $A=7.0$, about the equilibrium value $x_{{\rm s}0}=12$ and the time period being $T=2000 \tg$. In this study we have taken the frequency and ampltiude of oscillation as free parameters. As $\mdotsk$ decreases, $\xs$ moves outward and the net disk luminosity decreases resulting in a weaker radiation field (compare blue and red curves in Figs. \ref{fig:Shockcolsn}c$_1$-c$_5$). The  magnitude of net radiative force term $|R^r|$ for $\xs>x_{{\rm s}0}$ is smaller than the steady-state values (blue) and is larger for $\xs < x_{{\rm s}0}$. So in the funnel region, $v$ is higher when $\xs > x_{{\rm s}0}$ but eventually produces a jet whose $\vt$ is weaker and vice versa. Figures \ref{fig:Shockcolsn} a$_2$, b$_2$ and c$_2$, show the solution for the configuration when the accretion disk is brightest and the flow is suppressed to remain subsonic within the funnel. However, the flow velocity is boosted in the range $20<r<40$. The accelerated jet material encounters the slowly moving outer domain ($r>40$). As the supersonic fluid is resisted by relatively slower fluid ahead, it drives a shock in the jet, shown in  Fig. \ref{fig:Shockcolsn}($\rm a_3$). The formation of shock further reduces the downstream velocity of the fluid, as a result we can see the formation of multiple shocks in the flow namely ${\rm S_1,\,S_2,\,S_3}$ as shown in Fig.  \ref{fig:Shockcolsn}($\rm a_4$). The shock $\rm S_1$ being faster, catches up and collides with the shock in front of it ($\rm S_2$), shown in Fig. \ref{fig:Shockcolsn}($\rm a_5$). In Fig. \ref{fig:collision} we have plotted the zoomed region of jet domain $300<r<800$ to highlight the changes in mass density ($\rho$) distribution due to collision of the shocks. Panels
 \ref{fig:collision}$\rm(e_1)-(e_{5})$ show two shocks S$_1$ and S$_2$ collide and are plotted in the epoch between Fig.\ref{fig:Shockcolsn}$\rm (a_4)-(a_5)$. Figures \ref{fig:collision}$\rm(e_6)-(e_{10})$ clearly shows that the collision results in a cascade effect and induces additional shocks in the flow. Of course, if the oscillation is stopped, the jet eventually returns back to its steady-state.   

\begin{figure}[!h]
\centerline {\includegraphics[width=14cm, height=7cm]{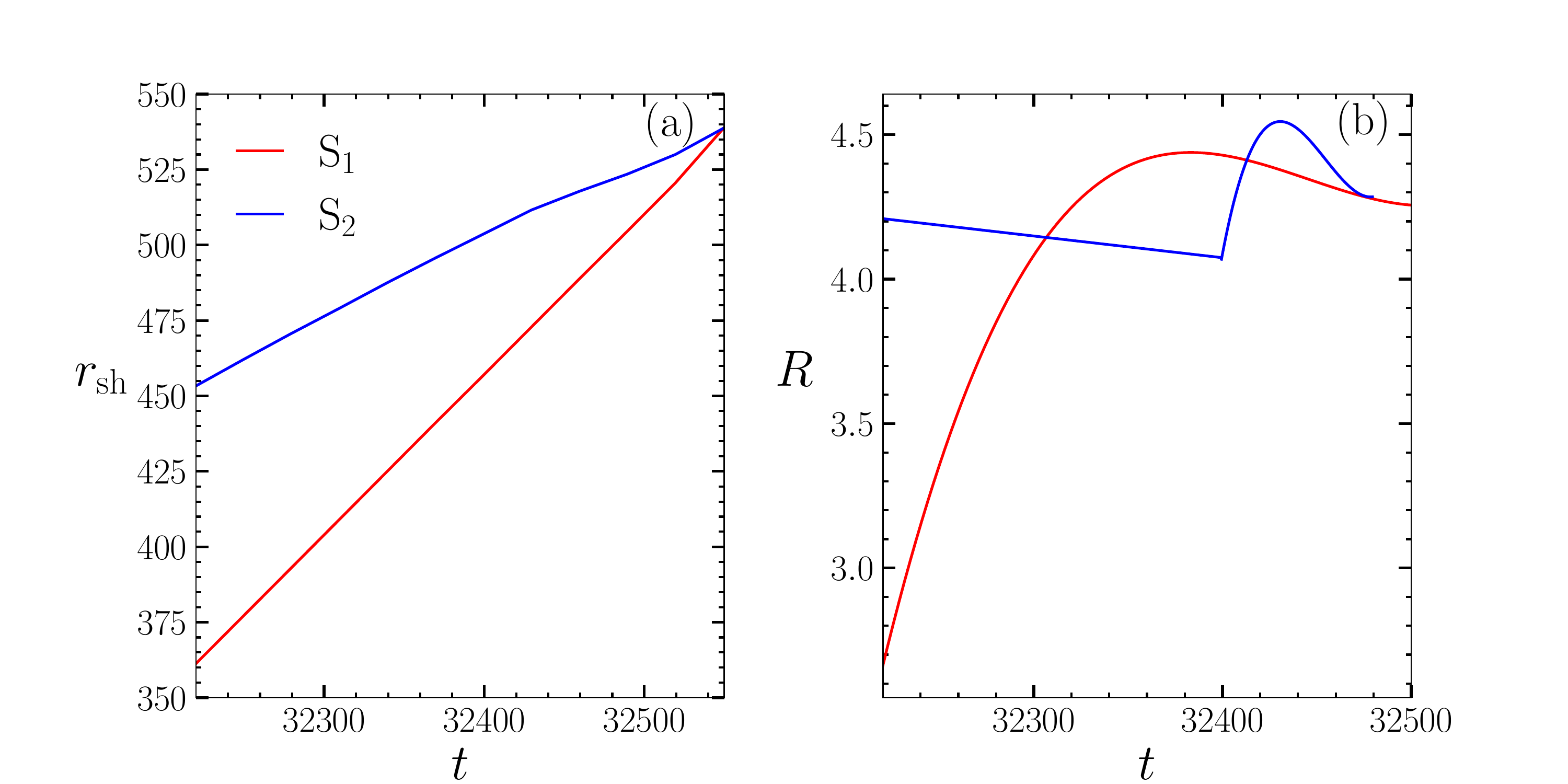}} 
\caption{The variation of shock positions and the compression ratio $R$ is plotted in panels (a) and (b), respectively.
Jets as described in Figs. \ref{fig:Shockcolsn} \& \ref{fig:collision}.}
\label{fig:pos_ratio}
\end{figure}

In Fig. \ref{fig:pos_ratio} (a), (b) we plot the variation of positions of jet shocks $\rm S_1$ and $\rm S_2$ and their compression ratios $R$ as a function of time $t$. Figure \ref{fig:pos_ratio} (a) shows that the slope of $\rsh$ vs $t$ graph is steeper for the shock $\rm S_1$, which clearly suggests that the speed of $\rm S_1$ is faster than that of $\rm S_2$ and ultimately catches up with $\rm S_2$ at $\rsh \sim 535$. The compression ratio is the ratio of post-shock and pre-shock densities ($R=\rho^+/\rho^-$) which is an indicator for the strength of the shock. Panel(b) shows that both the shocks are very strong but the variation of $R$ with respect to time has a complicated profile for both shocks. The shock $\rm S_1$ becomes stronger as it moves outwards, reaching a maximum value of $R\sim4.4$ and gradually starts to slow down. The compression ratio for second shock $\rm S_2$ reduces initially but starts to increase to reach a maximum value of $R\sim4.55$ and then it becomes weaker.  \\

\begin{figure}[!h]
\centerline {\includegraphics[width=22cm, height=9cm]{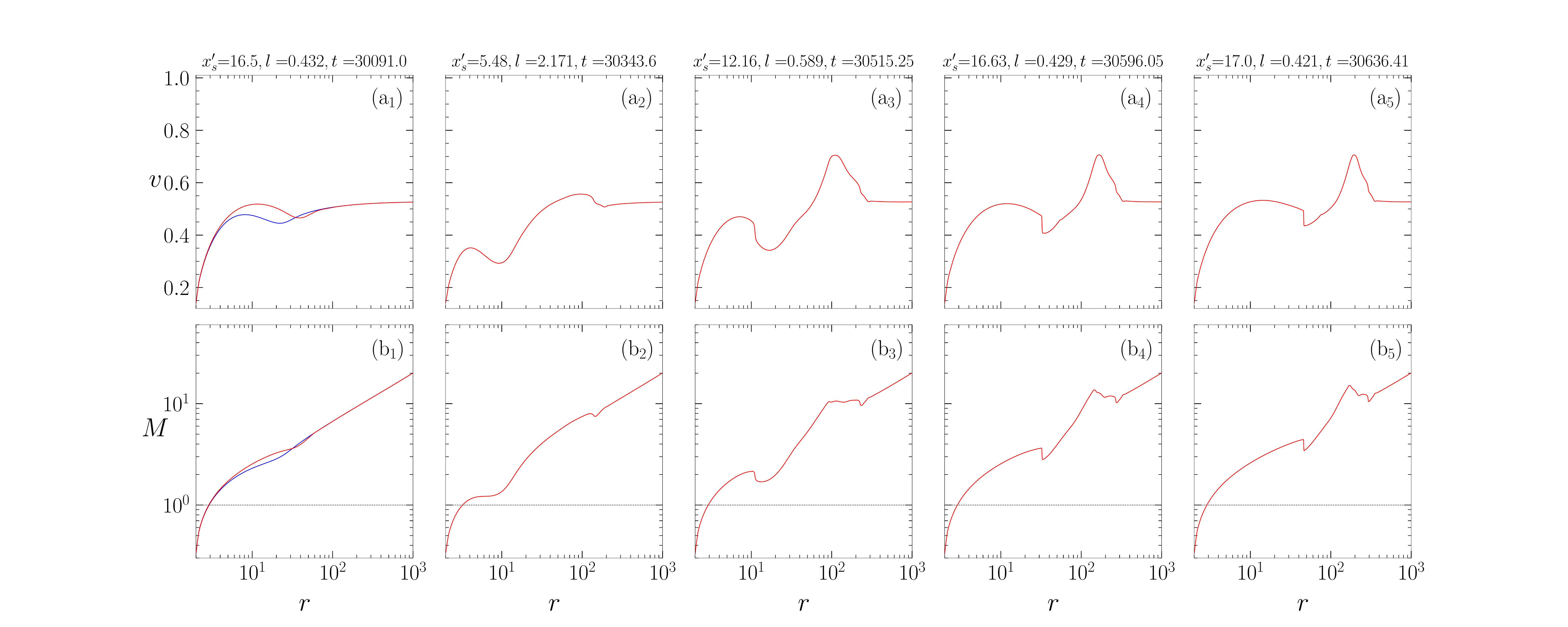}} 
\caption{The time dependent jet solution corresponding to injection parameters $\vin=0.14,\,\,\,\thetin=0.226,\,\,\,\rin=2.0$. }
\label{fig:int_shock}
\end{figure}

A closer look at equation \ref{eq:rad_mom_deps} indicates that the radiative effects are strong for the colder flows and relatively weaker for the hotter jets. Hence, we also study a hotter jet solution with injection parameters
$\vin=0.14,\,\,\,\thetin=0.226,\,\,\,\rin=2.0$, considered from the analytical jet solution for a steady disk configuration $\xs=11.02$ and $l=0.664$. The velocity ($v$) and Mach number ($M$) variation for the steady-state solution is plotted in Figs. \ref{fig:int_shock}($\rm a_1$) and ($\rm b_1$) with the solid blue line. The amplitude of oscillation for $\xs$ is $A=6.0$ and the time period of oscillation is $T=500 \tg$. In Figs. \ref{fig:int_shock}a$_1$---b$_5$ we have plotted jet solution for different values of $\xs$ as $\mdotsk$ is varied with time. So the variation of $\mdotsk$ determines the $l$ and $\xs$ at various $t$. One of the distinct feature of this hotter solution is that it harbors an internal shock very close to the jet base ($\rsh \sim10$) as shown in Fig. \ref{fig:int_shock}(a$_3$, b$_3$) and this shock is transported outwards as the jet solution evolves in time.
Since this is a hot jet near the base, radiative force is less effective, and it would be thermally accelerated to supersonic speeds within the first few Schwarzschild radii and more importantly within the funnel. A supersonic jet has low temperature and radiative terms become significant. Inside the funnel $R^r<0$, so resistance to the supersonic part of the jet drives a shock. This type of shock was not present in the previously shown solution because the jet base was relatively colder and so the jet was strongly decelerated to remain subsonic throughout the funnel region just above the PSD. 
It may be noted that such shocks in jets have been invoked to explain high energy emissions \citep{jfk08,lrwc11}.

\subsection{Effect of disk oscillation frequency on jet solutions}
\begin{figure}[!h]
\centerline {\includegraphics[width=22cm, height=12cm]{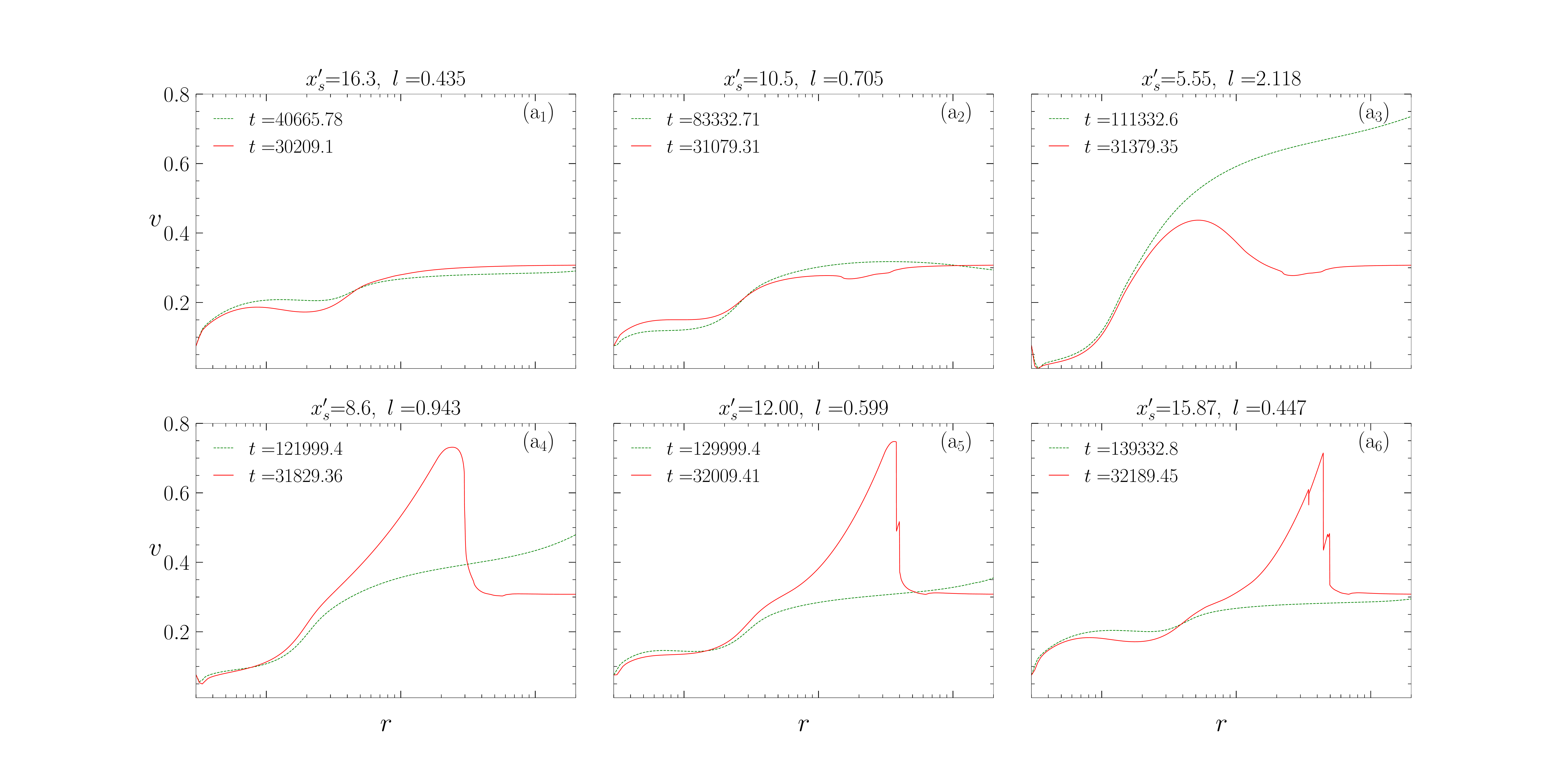}} 
\caption{The evolution of jet solutions ($v$ vs $r$) under the influence of disk oscillations of frequencies $f_s=2000\tg$ (solid, red)
and $10^5\tg$ (dashed, green). The injection and disk parameters are similar to Fig.(\ref{fig:Shockcolsn}). The two jet solutions in each panel corresponds to same $\xs$ and $l$, and the time in which the disk configuration is similar for the two cases are mentioned in the panels}.
\label{fig:freq_effect}
\end{figure}

We have discussed earlier in section \ref{sec:disc_inertial} that the information regarding the time variability of the disk (or portions of it) does not reach the entire jet simultaneously. The photons take some finite amount of time to reach different parts of the jet. Hence, the different points on the jet axis at a time $t$ would see the outer edge of the PSD at different locations. If the oscillation frequency of the disk is very low then the emitted photons can reach to a larger distance of the jet axis before the outer edge of the PSD reaches a new position. 
In Fig.(\ref{fig:freq_effect}) we have compared the jet solutions corresponding to two different frequencies of the disk oscillation. The injection parameters are $\vin=0.076,\,\,\,\thetin=0.064,\,\,\,\rin=3.0$ (same as Fig. \ref{fig:Shockcolsn}). The time period of the disk oscillation is $T=2000\,t_g$ for solution plotted with solid-red line and $T=10^5\,t_g$ for the solution plotted with dashed-green line. Here the jets are compared at the same disk configuration (same $l$ and $\xs$). The time at which the disk reaches same $l$ and $\xs$, will be different for the two frequencies, and has been shown on each panel. Assuming an $M_{\rm B}=10\,M_{\odot}$ these time periods correspond to oscillation frequencies 5 Hz and 100 mHz, respectively. The microquasars show the quasi periodic oscillations (QPOs) at various frequencies. For exmaple GRO J1655-40 (with central mass $\sim 7 M_{\odot})$ itself shows the QPOs at various frequencies extending from $0.1-10 {\rm Hz}$ \citep{cdp08,rmmbo98}. Hence, as a representative case for two different frequencies of disk oscillations we have taken the values $100\,{\rm mHz}$ and $5\,{\rm Hz}$. These results clearly indicate that for a low frequency oscillations in disk the transient features (shocks) are not present in jet solutions. The jet velocity gets boosted up or slowed down depending on the disk configuration but the solutions remain smooth throughout the domain.  

\section{Discussion and Conclusion}
\label{sec:concl}
In this paper, we studied the relativistic jets driven by the radiation field of advective type disks. The jet gains momentum supplied by the radiation field. We use relativistic radiation hydrodynamic equations along with a relativistic EoS to study the jet properties. We use the steady-state analytical solutions to verify our time-dependent code. The steady-state solutions also provide us the injection parameters at the jet base to be used in time-dependent cases. The accretion disk plays an auxiliary role in our study by supplying only radiation. It may be noted that we have considered sub as well as super-critical accretion rates to compute the radiation field above the disk. In the advective disk regime, super critical accretion rates may make the inner part of accretion disk optically slim ($1<$optical depth$\lsim$few), although higher infall velocity keeps the outer part of the disk optically thin. In such cases,
radiative transfer equation need to be solved to compute the radiative intensity emitted by the disk, especially from the inner part. Since the disk plays an auxiliary role and is not part of the computational domain, we considered simplifying assumptions to compute the radiation field. We show the effect of radiative acceleration on the terminal speeds of the jets. The super-critical disks accelerate the electron-proton jets to terminal Lorentz factor $\gamma_{\rm T}\gsim 4$. And for the jets driven by sub-critical disks also achieve reasonable relativistic speeds. The EoS allowed us to study the solutions corresponding to various plasma compositions. We found out that the electron-positron jets can be accelerated to ultra-relativistic speeds ($\gamma_{\rm T}>50$). The radiative acceleration is quite impressive, however, we have not considered the effect of back scattered photons onto the jet base, which may inhibit the jets. That might slightly lower such impressive terminal speeds we have obtained. In a nutshell, it may be summarized that the net radiation force ($R^r$) is negative inside the funnel region above the PSD and therefore opposes the jet expansion, but above the funnel region the net radiative force is positive and therefore the jet is accelerated. The time dependence of the jet is imposed by the time dependence of the radiation field. It may be recalled that \cite{kcm14} studied the radiatively and thermally driven outflows from advective accretion disks and they have shown that the radiative momentum deposition has a more significant effect on jet acceleration than conversion of thermal energy to the kinetic one. Moreover, they also found out that for a given $r_{\rm in}$ (jet base), $v_{\rm in}$ is higher for higher luminosity, but difference in sound speed or $\Theta_{\rm in}$ is imperceptible. Hence we have fixed the jet base for this study. It may be noted that, we have checked that by imposing a 20\% fluctuation of $\vin$ with a frequency similar to that of the oscillating $\xs$, it produces imperceptible change in the jet solution, therefore keeping the jet base values fixed was reasonable.
From its steady-state value as $\xs$ decreases, $l$ increases, the funnel becomes smaller, so there is larger acceleration at a smaller $r$, this faster region of the jet eventually is opposed by the slower moving outer jet, which eventually develops into a shock. 
Therefore an oscillating radiation field in the funnel like region above PSD produces shocks in the jet. The cold jet solution does not harbor a shock very close to the jet base, but jets with higher initial temperatures can have an internal shock very close to the jet base ($r_s\sim 10$).  
We have shown that the shocks can collide with each other and create multiple shocks after collision giving rise to shock cascade. We also showed that the production of shocks in jet can also depend on the frequency of oscillation of $\xs$ or $\mdotsk$. Oscillations with higher
frequency are likely to produce jet shocks compared to oscillations with lower frequency. In this work, we showed how the radiation field of the accretion disk can affect a jet in a complicated manner. The radiation field of the disk can produce mildly relativistic to highly relativistic jets
and we have also shown that disk oscillations give rise to the production and 
collision of shocks in jets. 

\section*{acknowledgments}
 We thank the anonymous referee for valuable comments which improved the quality of the paper. 

\appendix

\section{Sub-Keplerian accretion rate and the size of PSD}

The size of the PSD is related to the sub-Keplerian accretion rate \citep{vkmc15}
\begin{equation}
\xs=64.8735-14.1476\mdotsk+1.242\mdotsk^2-0.0394\mdotsk^3
\label{eq:xs_mdot}
\end{equation}

We give a perturbation in $\mdotsk$ given as 
\begin{equation}
\mdotsk=\dot{m}_{sk_0}-A_1sin(2\pi f_st)+A_2sin^2(2\pi f_s t)
\label{eq:accretvar}
\end{equation}
which results in  a sinusoidal oscillation in shock position with a frequency $f_s$.
$\dot{m}_{sk_0}$ is the mean value of accretion rate and $A_1$ and $A_2$ control the amplitude of oscillation in $\mdotsk$. 
For example, 
$\dot{m}_{sk_0}=7.5$, $A_1=3.6872$ and $A_2=1.313$, results in a sinusoidal oscillation in shock position with mean shock position $\xs=12$ and amplitude of oscillation is $A=7$ .

\section{Luminosity of disk}
The luminosities of disk components are obtained by numerically integrating the emissivity over the disk volume.

\begin{equation}
l_{i}=4\pi\int\mathcal{X}(\xs)I_{i}xhdx/L_{\rm edd}
\label{eq:lum1}
\end{equation}

Where $i$ represents the disk component ($i$=SKD or PSD), and $l_i$ is the luminosity in units of Eddington luminosity. $I_i$ is the Synchrotron emissivity ($\rm{erg\,cm^{-3}\,s^{-1}}$), $\mathcal{X}(\xs)$ is the Comptonization parameter fitting function. $\mathcal{X}(\xs)=1$ for SKD and for PSD it given as \citep{kcm14}

\begin{equation}
\mathcal{X}(\xs)=0.659234+0.127851\xs-0.00043\xs^2-1.13\times10^{-6}\xs^3
\label{eq:compt_factor}
\end{equation}
  
The disk luminosities are approximated by the algebraic functions given as      

\begin{equation}
l_{\rm SKD}=(93.88-0.12\xs^2)\xs^{-2.84}
\label{eq:skd_lum_func}
\end{equation}

\begin{equation}
l_{\rm PSD}=(55.89+1.145\xs^2)\xs^{-2.43}
\label{eq:psd_lum_func}
\end{equation}

\section{Numerically computed radiative moments from SKD} 
\label{sec:app1}
The radiative moments with relativistic transformations on the jet axis due to SKD are given as \citep{vkmc15}. 
It may be noted that, 
\begin{equation}
 {\cal E}={\cal E}_{\rm sk}+{\cal E}_{\rm ps};~{\cal F}={\cal F}_{\rm sk}+{\cal F}_{\rm ps};~\&~{\cal P}={\cal P}_{\rm sk}+{\cal P}_{\rm ps}
 \label{eq:num_mom}
\end{equation}
In the following we present the expressions of radiative moments computed from the SKD (e. g., $Q_{\rm sk}$) and PSD (e. g., $Q_{\rm ps}$). One need to follow Fig. (\ref{fig:cartoon}b) to get a proper reference of the following equations,
\begin{equation}
\mathcal{E}_{\rm sk}=\int_{x_i}^{\xo}\int_0^{2\pi} F_{\rm sk}\frac{ r dx d\phi}{\left[\left(r-x\rm{cot}\theta_{\rm sk}\right)^2+x^2\right]^{3/2} \gamma_{\rm sk}^4(1+\vartheta_il^i)^4_{\rm sk}},
\label{eq:skd_ez}
\end{equation}
\begin{equation}
\mathcal{F}_{\rm sk}=\int_{x_i}^{\xo}\int_0^{2\pi} F_{\rm sk}\frac{\left(r-x\rm{cot}\theta_{\rm sk}\right) r dx d\phi}{\left[\left(r-x\rm{cot}\theta_{\rm sk}\right)^2+x^2\right]^{2}\gamma_{\rm sk}^4(1+\vartheta_il^i)^4_{\rm sk}},
\label{eq:skd_fz}
\end{equation}
\begin{equation}
\mathcal{P}_{\rm sk}=\int_{x_i}^{\xo}\int_0^{2\pi} F_{\rm sk}\frac{\left(r-x\rm{cot}\theta_{sk}\right)^2 r dx d\phi}{\left[\left(r-x\rm{cot}\theta_{\rm sk}\right)^2+x^2\right]^{5/2}\gamma_{\rm sk}^4(1+\vartheta_il^i)^4_{\rm sk}},
\label{eq:skd_pz}
\end{equation}
where 
\begin{equation}
F_{\rm sk}=\mathcal{S}_{\rm sk}\left(\frac{\vartheta_{\rm o}\xo H_{\rm o}}{\vartheta_{\rm sk}xH} \right)^{3(\Gamma-1)} \frac{\left(x{\rm cos}\theta_{\rm sk}+d_0{\rm sin}\theta_{\rm sk}\right)}{xu\vartheta^2_{\rm sk}\left(x\rm{cot}\theta_{\rm sk}+d_0\right)^2} 
\label{eq:func_sk}
\end{equation}

where $\mathcal{S}_{\rm sk}$ is a constant given as
$\mathcal{S}_{\rm sk}=({9.22\times10^{33}e^4\Theta_0^3\beta\sigmat\mdotsk^2})/({\pi \mel^2\mp ^2c^2G^2\msol^2})$.
Here, $\gamma_{\rm sk}$ is the Lorentz factor, $\vartheta^i$ is the i-th contra variant component of three-velocity of accreting matter and $l^i$s are direction cosines. The temperature at outer edge of SKD $\Theta_{\rm 0}$ is taken to be $\Theta_{\rm 0}=0.2$. $\mdotsk$ represents the accretion rate of SKD in units of Eddington accretion rate $\left(\dot{M}_{\rm Edd}=1.44\times10^{17}M_{\rm{B}}/\msol \, \rm{gs^{-1}}\right)$. $\vartheta_{\rm sk}$ is the radial velocity of the accretion disk estimated from free fall velocity \citep[see][]{vkmc15}. The specific angular momentum $\lambda_0$ at outer edge is taken to be $\lambda_0=1.7$.\\

The specific intensity of PSD \citep{cc02} is given by
$I_{\rm{ps}}=l_{\rm ps}L_{\rm Edd }/\pi A_{\rm ps}$,
where $A_{\rm ps}$ and $L_{\rm Edd }$ are surface area of PSD and Eddington luminosity, respectively and luminosity of PSD $l_{\rm ps}$ is in units of $L_{\rm Edd }$.
The exact form of radiative moments from PSD are
\begin{equation}
\mathcal{E}_{\rm{ps}}=\mathcal{S}\int_{\xin}^{\xs}\int_0^{2\pi} \frac{ rxdx d\phi}{\left[\left(r-x{\rm cot}\theta_{\rm ps}\right)^2+x^2\right]^{3/2}\gamma_{\rm ps}^4(1+\vartheta_il^i)^4_{\rm ps}}
\label{eq:psd_ez}
\end{equation}

\begin{equation}
\mathcal{F}_{\rm{ps}}=\mathcal{S}\int_{\xin}^{\xo}\int_0^{2\pi} \frac{ r\left(r-x{\rm cot}\theta_{\rm ps}\right)xdx d\phi}{\left[\left(r-x{\rm cot}\theta_{\rm ps}\right)^2+x^2\right]^{2}\gamma_{\rm ps}^4(1+\vartheta_il^i)^4_{\rm ps}}
\label{eq:psd_fz}
\end{equation}

\begin{equation}
\mathcal{P}_{\rm{ps}}=\mathcal{S}\int_{\xin}^{\xs}\int_0^{2\pi} \frac{r\left(r-x{\rm cot}\theta_{\rm ps}\right)^2xdx d\phi}{\left[\left(r-x{\rm cot}\theta_{\rm ps}\right)^2+x^2\right]^{5/2}\gamma_{\rm ps}^4(1+\vartheta_il^i)^4_{\rm ps}}
\label{eq:psd_pz}
\end{equation}
Where $\mathcal{S}$ is a constant given as
$\mathcal{S}=({1.3\times10^{38}\lpsd\sigmat})/({2\pi\mel A_{\rm{ps}}G\msol})$.
Equations (\ref{eq:skd_ez}-\ref{eq:psd_pz}) are computed numerically.

\section{Analytical estimation of radiative moments}
\label{sec:app2}
In simulation code, we use the analytical estimates of radiative moments (mentioned in \ref{eq:num_mom}), the explicit expressions for these analytical functions are given below.\\   

The functions which mimic the radiative moments from SKD are given as 
\begin{equation}
\mathcal{E}_{\rm sk}=\mathcal{S}_{sk}\mathcal{C}_1(\xs)\left[E_f(r, x_o)-E_f(r, x_i)\right], \mbox{where}\,\,E_f(r, x)=-\frac{E_1(r,x)+E_2(r,x)}{E_3(r, x)}      
\label{eq:eng_alb}
\end{equation}

\begin{eqnarray*}
E_1(r,x)=rA(r,x)^{1/2}{\rm{sinh}}^{-1}\left(\frac{r^2-xr\cot\thsk}{|x||r|}\right) \nonumber \\
E_2(r,x)=\left[r\left(\cot^2\thsk-1\right)-\left(\cot^3\thsk+\cot\thsk\right)x\right]|r| ,\,\,\, E_3(r,x)=|r|A(r,x)^{1/2}r^2 \nonumber \\
\end{eqnarray*}

\begin{equation}
\mathcal{F}_{\rm sk}=\mathcal{S}_{sk}\mathcal{C}_2(\xs)\left[F_f(r, x_o)-F_f(r, x_i)\right]    
\label{eq:forc_alb}
\end{equation}

where
\begin{equation}
F_f(r, x)=-r\left[(F_1(r, x)+F_2(r,x)+F_3(r, x)+F_4(r,x)\right] \nonumber   
\end{equation}

\begin{eqnarray*}
F_1(r,x)=-\frac{{\rm{log}}[A(r,x)]}{2r^3},\,\,\,F_2(r,x)=\frac{1}{r^3}\cot\thsk\,{\rm{tan}}^{-1}\left[\frac{(1+\cot^2\thsk)x-r\cot\thsk}{r}\right],\,\,\, 
F_3(r,x)=\frac{1}{2rA(r,x)},\,\,\, F_4(r,x)=\frac{\rm{log}(x)}{r^3}  \nonumber \\ 
\end{eqnarray*}

\begin{equation}
\mathcal{P}_{\rm sk}=\mathcal{S}_{sk}\mathcal{C}_2(\xs)\left[P_f(r, x_o)-P_f(r, x_i)\right],\,\,\,P_f(r, x)=-\frac{P_1(r,x)+P_2(r,x)}{P_3(r, x)}    
\label{eq:pres_alb}
\end{equation}

\begin{eqnarray*}
P_1(r,r)=3zA(r,x)^{3/2} {\rm{sinh}}^{-1}\left[\frac{r^2-xr\cot\thsk}{xr}\right],\,\,\,P_3(r,x)=3r^2A(r,x)^{3/2}|r| \nonumber \\
P_2(r,x)=|r|[3x^2r(\cot^3\thsk-1)-(\cot^5\thsk+2\cot^3\thsk+\cot\thsk)x^3
+r^3(\cot^2\thsk-4)+(6\cot\thsk-3\cot^3\thsk)xr^2] \nonumber \\
\end{eqnarray*}

Function $A(r,x)$ is given as
\begin{equation}
A(r, x)=\left(r-x\,\cot\thsk\right)^2+x^2 \nonumber   
\end{equation}
and factors $\mathcal{C}_i(h_s)$ (i=1, 2, 3) are given as  
\begin{equation}
\mathcal{C}_1(\xs)=-410.9531+92629.4569/(1.0+(\xs/2.791155)^{2.22}))    
\end{equation}

\begin{equation}
\mathcal{C}_2(\xs)=\mathcal{C}_3(\xs)=252.7666+221.69\times10^8/(1.0+(\xs/0.001432)^{1.79}))    
\end{equation}
\\

The algebraic functions which mimic radiative moments from PSD are given as

\begin{equation}
\mathcal{E}_{\rm ps}=\mathcal{S}\mathcal{D}_1(\xs)\left[E_{ps}(r, \xs)-E_{ps}(r, \xin)\right], \mbox{where}\,\,E_{ps}(r, x)=-\frac{E_1(r,x)+E_2(r,x)\sqrt{A_{\rm ps}(r,x)}}{E_3(r,x)}      
\label{eq:eng_alb_ps}
\end{equation}

$E_1(r,x)=r^{\xs/e_1}{\rm sinh}^{-1}\left(\frac{r^2-rx\cot\thps}{|r||x|}\right)\left[B_1^2(\xs)r^3-2 xr^2B_1^2(\xs)\cot\thps + x^2r(B_1^2(\xs)\cot^2\thps+B_1^2(\xs))\right]$ \\

$E_2(r,x)=r^{\xs/e_2}\left[r^3-r^2\cot\thps(x-2B_1(\xs))+r(2xB_1(\xs)\cosec^2\thps+B_1^2(\xs)\cot^2\thps-B_1^2(\xs))-
xB_1^2(\xs)\cot\thps\cosec^2\thps\right]$ \\

$E_3(r,x)=(r^5-2xr^4\cot\thps+x^2r^3\cosec^2\thps)|r|$ \\
\\
where
$\left\{ 
  \begin{array}{ c l }
    \mathcal{D}_1(\xs)=96.936-420.455\,{\rm exp}(-0.609\xs),\,B_1(\xs)=1.40-2.182\,{\rm exp}(-0.44\xs),\,e_1=35.0, e_2=60.0 & \quad \textrm{if } \xs \leq 10 \\
    \mathcal{D}_1(\xs)=-0.073\xs^2+102.52,\,B_1(\xs)=1.58-0.48\,{\rm exp}(-0.10\xs),\,e_1=40.0,\,e_2=60.0                 & \quad \textrm{otherwise}
  \end{array}
\right.$ \\
\\

\begin{equation}
\mathcal{F}_{\rm ps}=\mathcal{S}\mathcal{D}_2(\xs)\left[F_{ps}(r, \xs)-F_{ps}(r, \xin)\right], \mbox{where}\,\,F_{ps}(r, x)=F_1(r,x)+F_2(r,x)+F_3(r,x)      
\label{eq:forc_alb_ps}
\end{equation}

\begin{eqnarray*}
F_1(r,x)=r^{(\xs/f_1)}B_2^2(\xs)\frac{log[A_{\rm ps}(r,x)]}{2r^3},\,\,\,F_2(r,x)=r^{(\xs/f_2)}\frac{\left[B_2(\xs)r-B_2^2(\xs)\cot\thps\right]}{r^3}\,{\rm tan}^{-1}\left[{\frac{(\cot^2\thps+1)x-r\cot\thps}{r}}\right] \nonumber \\
F_3(r,x)=r^{(\xs/f_3)}\left[\frac{-r^2+2x(r\cot\thps-B_2(\xs)\cot^2\thps-2B_2(\xs))+B_2^2(\xs)\cot^2\thps+B_2^2(\xs)}{2r^3(\cot^2\thps+1)-4xr^2(\cot^3\thps-\cot\thps)+2x^2r(\cot^4\thps+2\cot^2\thps+1)}+\frac{B_2^2(\xs)log(x)}{r^3}\right]
\end{eqnarray*}\\

where
$\left\{ 
  \begin{array}{ c l }
    \mathcal{D}_2(\xs)=169.13-145.21\,{\rm exp}(-0.116\xs),\,\,B_2(\xs)=0.05\xs+0.95,\,\,f_1=f_2=f_3=100.0 & \quad \textrm{if } \xs \leq 10 \\
    \mathcal{D}_2(\xs)=-3.05\xs+111.745,\,\,B_2(\xs)=1.4,\,\,f_1=35.0,\,f_2=32.0,\,f_3=40.0                 & \quad \textrm{otherwise}
  \end{array}
\right.$
\\
\\

\begin{equation}
\mathcal{P}_{\rm ps}=\mathcal{S}\mathcal{D}_3(\xs)\left[P_{ps}(r, \xs)-P_{ps}(r, \xin)\right],\,\,\,\,P_{ps}(r, x)=-\frac{P_1(r,x)+P_2(r,x)}{P_3(r,x)}      
\label{eq:pres_alb_ps}
\end{equation}
\\
Functions $P_1(r,x)$, $P_2(r,x)$, and $P_3(r,x)$ are given as \\
 
$P_1(r,x)=r^{\xs/25.0}\left[P_{11}(r,x)+P_{12}(r,x)\right]P_{13}(r,x)$ \\

$P_{11}(r,x)=3B_3^2(\xs)r^5-12xr^4B_3^2(\xs)\cot\thps+x^2r^3[18B_3^2(\xs)\cot^2\thps+6B_3^2(\xs)],\,\,\,P_3(r,x)={\rm sinh^{-1}}\left(\frac{r^2-rx\cot\thps}{|r||x|}\right)$\\ 

$P_{12}(r,x)=[-12B_3^2(\xs)\cot^3\thps-12B_3^2(\xs)\cot\thps]x^3r^2+[3B_3^2(\xs)\cot^4\thps+6B_3^2(\xs)\cot^2\thps+3B_3^2(\xs)]rx^4$\\

$P_{13}(r,x)={\rm sinh^{-1}}\left(\frac{r^2-rx\cot\thps}{|r||x|}\right)$ \\

$P_2(r,x)=r^{\xs/35.0}\sqrt{A_{\rm ps}(r,x)}\left[P_{21}(r,x)+P_{22}(r,x)+P_{23}(r,x)+P_{24}(r,x)\right]|r|$\\

$P_{21}(r,x)=r^5-(3x\cot\thps+2B_3(\xs)\cot\thps)r^4+\left[3x^2\cot^2\thps+(6B_3(\xs)\cot^2\thps+6B_3(\xs))x+B_3^2(\xs)\cot^2\thps-4B_3^2(\xs)\right]r^3$\\

$P_{22}(r,x)=-\left[x^3\cot^3\thps+(6B_3(\xs)\cot^3\thps+12B_3(\xs)\cot\thps)x^2-(6B_3^2(\xs)\cot\thps-3B_3^2(\xs)\cot^3\thps)x\right]r^2$\\

$P_{23}(r,x)=\left[(2B_3(\xs)\cot^4\thps+6B_3(\xs)\cot^2\thps+4B_3(\xs))x^3+(3B_3^2(\xs)\cot^4\thps-3B_3^2(\xs))x^2\right]r$\\

$P_{24}(r,x)=-x^3B_3^2(\xs)(\cot^5\thps+2\cot^3\thps+\cot\thps)$\\

$P_3(r,x)=\left[3r^7-12xr^6\cot\thps+x^2r^5(18\cot^2\thps+6)-12x^3r^4\cot\thps\cosec^2\thps+x^4r^3(3\cot^4\thps+6\cot^2\thps+3)\right]|r|$\\

here, $\mathcal{D}_3(\xs)=-2.48\xs+3.0/xs+76.45$, and $B_3(\xs)=-0.022\xs-0.056/xs+1.311$

\bibliography{biblio}{}
\bibliographystyle{aasjournal}



\end{document}